\documentclass[10pt,preprint]{emulateapj}
\usepackage{CJK}

\newcommand{\kms}{\hbox{km s$^{-1}$}}

\shorttitle{The eruption of ASASSN-15qi}
\shortauthors{Herczeg et al.}

\begin{document}
\begin{CJK*}{UTF8}{gbsn}

% Title.
\title{THE ERUPTION OF THE CANDIDATE YOUNG STAR ASASSN-15QI}

% Authors.
\author{Gregory J. Herczeg (沈雷歌)\altaffilmark{1}, Subo Dong\altaffilmark{1},
  Benjamin J. Shappee\altaffilmark{2,3}, Ping Chen (陈
  平)\altaffilmark{1,4}, \\
Lynne
  A. Hillenbrand\altaffilmark{5}, Jessy Jose\altaffilmark{1}, 
  Christopher S. Kochanek\altaffilmark{6,7}, Jose L. 
  Prieto\altaffilmark{8,9},\\
 K.~Z.~Stanek\altaffilmark{6,7}, Kyle
  Kaplan\altaffilmark{10}, Thomas W.-S.
  Holoien\altaffilmark{6,7}, Steve Mairs\altaffilmark{11,12}, Doug
  Johnstone\altaffilmark{11,12}, \\
Michael
  Gully-Santiago\altaffilmark{1}, Zhaohuan Zhu\altaffilmark{13}, 
Martin C.  Smith\altaffilmark{14}, David
  Bersier\altaffilmark{15}, \\Gijs D. Mulders\altaffilmark{16,17},
Alexei V. Filippenko\altaffilmark{18}, Kazuya Ayani\altaffilmark{19}, Joseph Brimacombe\altaffilmark{20},\\
Jonathan S. Brown\altaffilmark{6,7},
Michael Connelley\altaffilmark{21},
  Jussi
  Harmanen\altaffilmark{22},
Ryosuke
  Itoh\altaffilmark{23,24},\\ Koji S. Kawabata\altaffilmark{24,23},
Hiroyuki Maehara\altaffilmark{25}, Koji
  Takata\altaffilmark{23,24}, Heechan Yuk\altaffilmark{18}, WeiKang Zheng\altaffilmark{18}}%\\
\altaffiltext{1}{Affilations are listed at the end of the paper;
  contact gherczeg1@gmail.com}

\begin{abstract}
Outbursts on young stars are usually interpreted as accretion bursts
caused by instabilities in the disk or the star-disk connection.
However, some protostellar outbursts may not fit into this framework.
In this paper, we analyze optical and near-infrared spectra and photometry
to characterize the 2015 outburst of the probable young star
ASASSN-15qi.  
The $\sim 3.5$ mag brightening in the $V$ band was sudden,
with an unresolved rise time of less than one day.  The outburst
decayed exponentially by 1 mag for 6 days and then
gradually back to the pre-outburst level after 200 days.  
The outburst is dominated by emission from $\sim10,000$ K gas.  An 
explosive release of energy accelerated matter from the
star in all directions, seen in a spectacular cool, spherical wind
with a maximum velocity of 1000 \kms.   The wind and hot gas both
disappeared as the outburst faded and 
the source returned to its quiescent F-star spectrum.
Nebulosity near the star brightened with a delay of 10--20 days.
Fluorescent excitation of H$_2$
 is detected in emission from vibrational levels as high as $v=11$,
 also with a possible time delay in flux increase.
The mid-infrared spectral energy distribution does not indicate
the presence of warm dust emission, although the optical photospheric
absorption and CO overtone emission could be related to a gaseous disk.
Archival photometry reveals a prior outburst in 1976.  Although we
speculate about possible causes for this outburst, none of the
explanations are compelling.
\end{abstract}
 
% Official ApJ keywords in alphabetical order.
% See http://www.journals.uchicago.edu/ApJ/information.html

%%%%%%%%%%%%%%%%%%%%%%%%%%%%%%%%%%%%%%%%%%%%%%%%%%%%%%%%%%%%%%

\section{INTRODUCTION}

Large brightness changes of young stars were seen long before
the class of objects was understood to be related to star formation, or even that stars formed
\citep{hind,ceraski06}.  The largest and most prominent physical changes
are decades-long bursts of accretion at
rates of  $10^{-4}-10^{-5}$ M$_\odot$ yr$^{-1}$, called FUor objects
\citep[after FU Ori; see reviews by][]{herbig77,hartmann96,reipurth10}, and months-long bursts of
accretion at 
$\sim10^{-7}$ M$_\odot$ yr$^{-1}$ \citep{aspin10,lorenzetti12}, 
called EXor outbursts \citep[after EX Lup;][]{herbig89}.  
The FUor and EXor classes of outbursts form our framework for interpreting
large luminosity increases of young stars \citep[see review by][]{hartmann16}.  

The different timescales for the longer (FUor) and shorter (EXor)
outbursts suggest that they are different phenomena.
FUor outbursts are thought to be triggered by instabilities in the disk
\citep[e.g.,][]{armitage01,vorobyov05,zhu09,bae14} at radii with rotational timescales of
decades.  On the other hand, the shorter EXor outbursts are thought to be triggered
by instabilities in the magnetic connection between the star and disk
\citep{dangelo10,dangelo12}.  These two phenomena require different
disk structures, which is supported with spectroscopic evidence.  Viscously heated,
optically thick FUor disks produce low-gravity spectra similar to those of
supergiant stars, including deep absorption in
CO overtone bands at 2.3 $\mu$m \citep[e.g.,][]{greene08}.  EXors show a forest of optical
emission lines, evidence of magnetospheric accretion, and a warm disk
surface layer that produces strong CO emission \citep[e.g.,][]{herbig89,aspin10,caratti13,holoien14,sicilia15,banzatti15}.

Since measuring the duration of an outburst usually requires impatient
people to wait, these spectroscopic proxies are used to immediately
discriminate between FUor and EXor outbursts.   In practice, outbursts of protostars are often forced into
the EXor/FUor classification scheme, even when the outburst does not fit well
into either category.  Many outbursts appear to be intermediate
between the FUor and EXor classes \citep[e.g.,][]{contreras16}, despite the very
different mechanisms and masses thought to be involved.  In other cases, some characteristics
of the outburst are inconsistent with this classification scheme
\citep{ninan15}.  Either the classification system groups together
diverse physics, or the same accretion burst physics of EXors produces a wider range of observed
phenomena than expected.

Powerful new transient surveys are now leading to the discovery of 1--2 large
outbursts of young stellar objects (YSOs) each year \citep[e.g.,][]{miller11,holoien14}.  
A recent outburst of a candidate young star, ASASSN-15qi, was 
identified by the All-Sky Automated Survey for Supernovae (ASAS-SN) variability survey
\citep{shappee14}.  
Follow-up observations were obtained because the object was suspected
to be young \citep[see discussion in][]{hillenbrand15} based on its projected spatial location
near the H II regions Sh 2-148 and Sh 2-149 and the small, low
extinction molecular cloud TGU 676/Dobashi 3359
\citep{dobashi05,dobashi11}.  This region is a few degrees southwest of the main
Cepheus star forming complexes, including LDN 1218
\citep[e.g.,][]{kun08,allen12}.     A distance of $3.24$ kpc is
adopted from the parallax measured for G108.47$-$2.81, which is visually
located 1.5 deg from ASASSN-15qi and shares the same radial velocity
\citep{choi14}.

Initial optical spectroscopy of ASASSN-15qi showed P Cygni profiles
characteristic of strong winds, an indicator
of accretion-ejection processes that occur on young stars
\citep{maehara15,hillenbrand15,connelley15}.  However, subsequent 
low resolution near-infrared (IR) and 
high resolution optical spectra revealed a confusing picture of an
outburst that
does not neatly fit into either the FUor or EXor category
\citep{hillenbrand15,connelley15}.  Spatially extended emission in
optical imaging
\citep{hillenbrand15,stecklum15b} confirmed that the object is likely
young.   The source quickly faded from its peak
\citep{stecklum15a,stecklum15b}.

In this paper, we analyze optical and near-IR photometry and spectroscopy
of the 2015 outburst of ASASSN-15qi.  
The outburst is especially remarkable because of (1) the dramatic 3.5 mag brightening
in less than one day, and (2) the presence of a fast wind that either caused
or was produced by the outburst.  The wind faded as the outburst decayed
over 3 months.
In \S 2, we describe the wide array of observations used in this
paper.  In \S 3, we analyze these data to arrive at some
empirical conclusions.  In \S 4, we summarize the source properties,
compare these properties to those of known outbursts where the physics
is better understood, and describe some possible alternatives.  These
descriptions and speculations are summarized in \S 5.

\section{OBSERVATIONS}

The outburst of ASASSN-15qi (2MASS J22560882+5831040) occurred on JD 2,457,298 (2 October 2015; UTC dates are used herein).
This section describes the follow-up photometry and spectroscopy
obtained for 3 months after the burst occurred.    Archival photometry is listed in Table~\ref{tab:photarchive}, while
photometry obtained or measured as part of this paper is listed in
Tables~\ref{tab:asassnphot}--\ref{tab:nirphot}.    The log of
spectroscopic observations of ASASSN-15qi is listed in Table~\ref{tab:speclog}.

\begin{table}[!h]
\caption{Archival Photometry$^a$}
\label{tab:photarchive}
\begin{tabular}{lcccc}
\hline
Telescope/Survey & Epoch & Band & Mag & Ref.\\
\hline
USNO-A2 & 1953.83 & $B$ & 18.2 & 1\\
USNO-B1 & 1976.5 & $B$ & 17.81 & 2\\
USNO-B1 & 1976.5 & $B$ & 17.41 & 2\\
USNO-B1 & 1976.5 &$R$ & 14.62$^b$ &2\\
USNO-B1 & 1976.5 & $R$ & 14.76$^b$ & 2\\
USNO-B1 & 1976.5 & $I$ & 14.22$^b$ &2\\
GSC & 1989.669 & $B$ & 18.51 & 3\\
GSC & 1989.669 & $Bj$ & 17.84 & 3\\
GSC & 1989.669 & $V$ & 16.50 & 3\\
GSC & 1989.669 & $N(I)$ & 14.33 & 3\\
2MASS & 2000.06.21.39 & $J$ & 13.70 &4\\
2MASS & 2000.06.21.39 & $H$ & 12.92 & 4\\
2MASS & 2000.06.21.39  & $K$ & 12.65 &4\\
PTF & 2009.07.31.4 & $g$ & 17.06 & 5\\
PTF &  2009.08.25.3 &  $g$&     17.79 & 5\\
PTF &    2009.12.17.1 .& $g$ &      18.01  & 5\\
IPHAS & 2004.08.25  & $r$ & 16.66 & 6\\
IPHAS & 2004.08.25  & $i$ & 15.59 & 6\\
IPHAS & 2004.08.25 & H$\alpha$ & 16.04 & 6\\
Tautenburg 1.34m & 2004.09.09.90 & $I$ & 15.08 & 7\\
\hline
Leicester 0.5m & 2010.10.09.96       & $I$   & $15.08$ & 8\\
Tautenburg 1.34m & 2015.10.11.81 & $B$ & $16.13$ & 8\\
Tautenburg 1.34m & 2015.10.11.81 & $V$ & $15.17$ & 8\\
Tautenburg 1.34m & 2015.10.11.81 &$ R$ & $14.22$ & 8 \\
Tautenburg 1.34m & 2015.10.11.81 & $I$  & $13.11$ & 8\\
Tautenburg 1.34m & 2015.11.05.77 & $B$ & $17.05$ & 7 \\
Tautenburg 1.34m & 2015.11.05.77 & $V$ & $15.79$ & 7\\
Tautenburg 1.34m & 2015.11.05.77 & $R$ & $14.94$ & 7 \\
Tautenburg 1.34m & 2015.11.05.77 & $I$  & $13.80$ & 7 \\
Tautenburg 1.34m & 2015.11.13.99 & $B$ & $17.16$ & 7 \\
Tautenburg 1.34m & 2015.11.13.99 & $V$ & $15.99$ &7 \\
Tautenburg 1.34m & 2015.11.13.99 &$ R$ & $15.23$ &7 \\
Tautenburg 1.34m & 2015.11.13.99 & $I$  & $13.92$ &7 \\
IRTF & 2015.10.15.44 & $J$ & $12.15$ & 9\\
IRTF & 2015.10.15.44 & $H$ & $11.62$ &9\\
IRTF & 2015.10.15.44 & $K$ & $11.29$ &9\\
\hline
\multicolumn{5}{l}{$^a$All photometry in mag}\\
\multicolumn{5}{l}{$^b$Likely epoch of past outburst}\\
\multicolumn{5}{l}{1:  \citet{monet98}}\\
\multicolumn{5}{l}{2:  \citet{monet03}}\\
\multicolumn{5}{l}{3:  GSC2.2, STScI 2001}\\
\multicolumn{5}{l}{4:  \citet{cutri03}}\\
\multicolumn{5}{l}{5: \citet{law09,laher14}}\\
\multicolumn{5}{l}{6: \citet{barentsen14}}\\
\multicolumn{5}{l}{7: \citet{stecklum15b}}\\
\multicolumn{5}{l}{8: \citet{stecklum15a}}\\
\multicolumn{5}{l}{9: \citet{connelley15}}\\
\end{tabular}
\end{table}

\subsection{ASAS-SN V-band Photometry}

The ASAS-SN is an all-sky $V$-band transient survey with a limiting magnitude of
$\sim 17$.  ASAS-SN photometry of ASASSN-15qi was
obtained from two different telescopes on Haleakala, Hawaii with intervals of 1--3 days.   
The outburst of ASASSN-15qi was first detected and 
reported in the ASAS-SN transient list \citep{shappee14}.  
ASAS-SN monitoring includes extensive photometry before and during the
outburst.

The quiescent brightness of ASASSN-15qi is near
the limiting magnitude of ASAS-SN.  As a consequence, 
flux measurements prior to outburst
required a different technique than the standard ASAS-SN
photometry available online.  We first measured the flux in a
reference image, and subsequently calculate the flux in difference
images.  The fluxes are then extracted with a 2-pixel ($15\farcs6$) aperture radius.
Small offsets may occur
between the two telescopes.  The ASAS-SN photometry includes the
central source and all nearby nebulosity because of the large pixels.
A selected subsample of ASAS-SN photometry is listed in
Table~\ref{tab:asassnphot}; all photometry is available online.

\begin{table}[!ht]
{\begin{center}
\caption{Subsample of ASAS-SN Photometry$^a$}
\label{tab:asassnphot}
\begin{tabular}{ccc}
\hline
 JD$-$2,457,000 & $V$ (mag) & $1\sigma$ Error\\
\hline
275.91551    &  17.37 & 0.30\\
277.96895   &     16.88 &  0.17\\
288.89590  &     17.22 &  0.29  \\
297.87394   &     17.37 &   0.34\\
297.88712   &    16.99 &  0.25\\
\hline
298.77584   &     13.53 &  0.02\\
298.78226   &    13.60 &  0.02\\
300.83682   &    13.84 &  0.02\\
304.95129   &     14.70 &  0.04\\
305.90914    &   14.73 &  0.04\\
309.81402   &    14.75 &  0.04\\
309.84820   &     14.81 &  0.04 \\
314.80541    &   15.03 &  0.05\\
317.87615   &     15.29 &  0.07 \\
319.69468   &     15.26 &  0.09\\
323.81185   &    15.36 &  0.09\\
324.84534   &     15.27 &   0.07 \\
327.85558   &    15.44 &  0.07\\
333.76965   &    15.52 &  0.07\\
342.82343   &     15.73 &   0.09\\
344.73217   &    15.93 &  0.09 \\
360.72276   &    15.90 &  0.08\\
361.79525    &    15.87 &  0.09\\
363.79191   &   15.93 &  0.09 \\
\hline
\multicolumn{3}{l}{$^a$Complete photometry is online.}
\end{tabular}\end{center}}
\end{table}

\begin{table*}[!ht]
{\begin{center}
\caption{Ground-based Optical Photometry}
\label{tab:photometry}
\begin{tabular}{cccccccccc}
\hline
Date & \multicolumn{4}{c}{$5^{\prime\prime}$ Aperture photometry} &
\multicolumn{4}{c}{Aperture-PSF photometry} & Seeing\\
JD--2,457,000& $B$ & $V$ & $r$ & $i$ & $\Delta B$ & $\Delta
V$ &$\Delta r^a$ &$\Delta i^a$ & ($^{\prime\prime}$)\\
\hline
\multicolumn{10}{c}{LCOGT Photometry}\\
\hline
     318.57 &   16.38 &   15.15 &   14.63  &   14.04 &    0.31 &    0.18 &    0.23  &    0.20 & 3.2\\
     322.57 &   16.70 &   15.46 &   14.96  &   14.30 &    0.49 &    0.41 &    0.28  &    0.28 & 2.8\\
     324.67 &   16.99 &   15.66 &   15.12  &   14.42 &    0.53 &    0.35 &    0.36  &    0.27 & 2.5\\
     330.67 &   16.95 &   15.63 &   15.08  &   14.40 &    0.52 &    0.31 &    0.32  &    0.28 & 2.1\\
     333.62 &   16.91 &   15.63 &   15.05  &   14.39 &    0.50 &    0.28 &    0.29  &    0.27 & 2.0\\
     336.67 &   17.18 &   15.90 &   15.26  &   14.56 &    0.54 &    0.45 &    0.42  &    0.37 & 1.8\\
     339.67 &   17.46 &   15.93 &   15.31  &   14.61 &    0.26 &    0.28 &    0.35  &    0.29 & 2.5\\
     351.61 &   17.50 &   16.12 &   15.54  &   14.79 &    0.36 &    0.29 &    0.29  &    0.26 & 2.4\\
     356.67 &   17.78 &   16.31 &   15.65  &   14.93 &    0.31 &    0.15 &    0.26  &    0.24 & 3.1\\
     356.73 &   17.66 &   16.18 &   15.63  &   14.91 &    0.53 &    0.27 &    0.25  &    0.24 & 2.9\\
     360.62 &   17.54 &   16.20 &   15.57  &   14.85 &    0.42 &    0.24 &    0.28  &    0.30 & 2.7\\
     365.70 &   17.67 &   16.33 &   15.72  &   14.98 &    0.58 &    0.34 &    0.29  &    0.26 & 2.6\\
     370.55 &   17.80 &   16.26 &   15.67  &   14.92 &    0.25 &    0.14 &    0.14  &    0.18 & 3.9\\
     374.65 &   17.89 &   16.24 &   15.65  &   14.90 &    0.21 &    0.23 &    0.20  &    0.22 & 2.4\\
     378.62 &   17.70 &   16.27 &   15.70  &   14.96 &   ...  &   0.23 &    0.20  &    0.23  & 2.3\\
     379.60 &   17.82 &   16.27 &   15.76  &   15.02 &    0.27 &    0.30 &    0.20  &    0.21 & 2.2\\
 \hline
\multicolumn{10}{c}{Brimacombe Photometry ($\sim$ V)$^a$}\\
\hline
    387.63 &  &   16.18 &  &  &  &    0.09 &  &     4.5 \\
    390.66 &  &   16.15 &  &  &  &    0.10 &  &     4.2 \\
    392.65 &  &   16.17 &  &  &  &    0.16 &  &     3.5 \\
    397.63 &  &   16.31 &  &  &  &    0.13 &  &     4.6 \\
    399.62 &  &   16.26 &  &  &  &    0.11 &  &     4.6 \\
    401.61 &  &   16.29 &  &  &  &    0.05 &  &     5.6 \\
    405.64 &  &   16.35 &  &  &  &    0.17 &  &     3.8 \\
    408.59 &  &   16.37 &  &  &  &    0.15 &  &     4.1 \\
    415.60 &  &   16.45 &  &  &  &    0.09 &  &     5.1 \\
    424.62 &  &   16.52 &  &  &  &    0.09 &  &     4.3 \\
    427.63 &  &   16.55 &  &  &  &    0.01 &  &     5.5 \\
554.90 & ... & 16.71 &(15.74) & (15.65) & .. & 0.07 & 0.10 & 0.06 & 3.5\\
\hline
\multicolumn{10}{c}{Liverpool Telescope Photometry}\\
\hline
504.69 & 18.86 & 17.29 & 16.51 & 15.68 & 0.34 & 0.25 & 0.24 & 0.13 & 1.5\\
551.71 & 18.99 & 17.31 & 16.52 & 15.66 & 0.18 & 0.21 & 0.20 & 0.13 & 1.3\\
\hline
\multicolumn{10}{l}{$^a$Approximately $V$-band}\\
\end{tabular}
\end{center}}
\end{table*}

\subsection{Ground-based Optical Photometry}

We obtained $BVri$ images with the Las Cumbres Observatory Global
Telescope Network \citep[LCOGT; ][]{brown13} 1-m telescope at McDonald
Observatory between 2015 October 23 and December 23 and at the
Liverpool Telescope  on 2016 April 26 and 2016 June 11.
Additional
optical photometry in approximately the $V$ band was obtained by Joseph Brimacombe in 2015 December
-- 2016 January from New Mexico Skies, New Mexico, USA using a 43-cm Planewave CDK telescope on a
Software Bisque PME II mount with an SBIG STL-6303 camera and
Astrodon Johnson-Cousins photometric filters.  
Results are presented
in Table~\ref{tab:photometry}. 
  
The astrometry for these images was obtained using Astrometry.net
\citep{lang10}.  The pixel scales are $0\farcs467$ for LCOGT,
$0\farcs30$ for the Liverpool Telescope, and $0\farcs63$ for the
Brimacombe photometry.

Relative source fluxes were measured in apertures
with a $5^{\prime\prime}$ radius using the 
{\it apphot} task in IRAF \citep{tody93}.  The absolute calibration was then 
calculated from the AAVSO Photometric All-Sky Survey
\citep{Henden15}.  Calibration uncertainties are typically 0.1 mag in $B$, 0.06 mag in
$V$, 0.05 mag in $r$, and 0.07 mag in $i$.  Random statistical errors
introduced by photon noise and other relative calibration
uncertainties are $0.03$ mag per band.
However, systematic uncertainties of 0.1--0.5 mag exist and are dominated by
methodological differences in how the nebulosity affects the flux
measurement (see examples for {\it Swift} data in Table~\ref{tab:swift}).
These uncertainties are minimal near the outburst peak but significant
when the central source is weak.  For the Liverpool imaging, obtained
207 days after outburst peak, point-spread function (PSF) photometry and photometry with
a $1\farcs5$ radius extraction region both yield brightnesses that are
$\sim 0.5$ mag fainter in all bands.

\begin{table*}[!t]
{\begin{center}
\caption{{\it Swift} Photometry}
\label{tab:swift}
\begin{tabular}{cccccccccc}
\hline
Date & Post-burst & Obsid & Measurement$^a$ & $UVW2$ & $UVM$ & $UVW1$ & $U$ & $B$ & $V$ \\
% \multicolumn{2}{c}{$\lambda_{mean}$} & 2080  & 2255 & 2614 & 3471 & 4359 & 5430 \\
\hline
\hline
12 Oct 2015 & 10 days& 00034098001 & $t_{\rm exp}$ (s) &  314.7 & 1231.9 &157.3 &78.5 & 78.5  & 78.5 \\
\hline
& && $5^{\prime\prime}$ aperture (mag) &19.12 & 20.70 & 17.97
&  16.85 & 16.24 & 14.94 \\
&  & & Gaussian fit (mag) &   19.67 & 20.73 & 18.41 & 16.95 & 16.25 & 15.12\\
\hline
\hline
27 Dec 2015& 87 days & 00034098002 & $t_{\rm exp}$ (s) & 1633.1 &--& 817.2  &407.8 & 451.3 &407.8 \\
\hline
& && $5^{\prime\prime}$  aperture (mag) & -- & -- & 20.17 &
18.84 & 18.10 & 16.53\\
&  & &  Gaussian fit (mag) & $>21.9^b$ & -- & 20.24 &
19.02 & 18.22 & 16.93\\
\hline
\multicolumn{9}{l}{$^a$Magnitudes measured either from aperture
  photometry with $5^{\prime\prime}$ extraction radius or Gaussian fits.}\\
\multicolumn{9}{l}{$^b$3$\sigma$ upper limit.}\\
\end{tabular}\end{center}}
\end{table*}

\begin{table}[!b]
{\begin{center}
\caption{Near-IR and Mid-IR Photometry}
\label{tab:nirphot}
\begin{tabular}{lccccc}
\hline
Telescope & JD--2,457,000 & $I1$ & $I2$ & $I3$ & $I4$\\
\hline
{\it Spitzer}/IRAC  &  $-2091.5$ &  12.36 & 12.30 & 12.50 & $>11.9$\\
\hline
Telescope & JD--2,457,000 & $J$ & $H$ & $Ks$\\
\hline
NOT     & 379.5 & 12.93 & 12.30 & 11.85 & \\
Liverpool &385.35 &... & 12.40 & ...\\
Liverpool & 392.88 &... & 12.56 & ...\\
Liverpool & 551.71 & ... & 12.76 & ...\\
\hline
\\
\end{tabular}\end{center}}
\end{table}

\begin{table*}[!t]
\caption{Spectroscopy of ASASSN-15qi}
\label{tab:speclog}
\begin{tabular}{ccccccccccc}
\hline
 Telescope & Instrument & Wavelength ($\mu$m) & Slit width ($^{\prime\prime}$) & Resolution &
$t_{\rm exp}$ (s) & JD$^a$
& Post-burst$^a$ & $V_{\rm est}$ \\
\hline
Kanata 1.5m & HOWPol & 0.45--0.90 & 2.2 & 400 &300 &  303.499 & 5.5 & 14.2\\
Bisei 1.01m  & ...  &  0.40--0.80 & 2.0 & 1300 & 2400 &  304.014 & 6.0 & 14.3\\
NASA/IRTF & SpeX & 0.7--2.5 & $0.8$ &1300 &1440 & 310.940 & 12.9 & 14.8\\
Lick/Shane & Kast &  0.35--1.0 & 2 & 600 & 900 &  315.5 & 17.5 & 15.0 \\
McDonald 2.7m& IGRINS & 1.44-2.5 & 1  & 40000 & 4800 & 318.599 & 20.6 & 15.1 \\
Keck-I & HIRES & 0.48--0.90    & 0.86   & 45000     &  1200 &  322.267 & 24.7 & 15.4\\
MDM     &   OSMOS & 0.43--0.68 & 1.2 & 1800 &1200 & 338.675 & 40.7 & 15.7\\
Keck-I & HIRES  & 0.48--0.90   &  1.15   &  45000 & 1800  & 377.226 & 79.7 & 16.3\\
Keck-I & MOSFIRE & 2.0--2.3  & 1.1 & 3000 &400 & 386.245 & 88.7 & 16.3\\
Liverpool   & SPRAT & 0.50--0.75   & $1\farcs8$ & 350 &500 & 392.395 & 94.4 &  16.3\\
Liverpool & SPRAT & 0.50--0.75   & $1\farcs8$ & 350 &1000 & 504.68 &206.7 & 17.1 \\
LBT  & MODS & 0.35--0.9 & $1\farcs2$ & $1800$ & $1800$ & 551.94 & 353.9 & 17.3\\
\hline
\multicolumn{9}{l}{$^a$JD--2,457,000; the burst occurred on JD 2,457,298.}\\
\multicolumn{9}{l}{$^bV$ at time of observation estimated from
  $V$-band lightcurve from $5^{\prime\prime}$ radius apertures.}\\
\end{tabular}
\end{table*}

\subsection{Swift Ultraviolet-Optical Photometry}

We triggered the {\it Swift} satellite to obtain images with the Ultraviolet/Optical Telescope
\citep[UVOT;][]{roming05} on 12 October 2015 (ID 00034098001) and 27
December 2015 (ID 00034098002).  The UVOT images have a field-of-view of
$\sim8^\prime$ with $0\farcs5$ pixels and were obtained with filters
covering 2000--6000 \AA.  Count rates were converted to fluxes and magnitudes
using the zero points calculated by \citet{poole08}.

ASASSN-15qi was detected in all filters on 12 October and in the $UW1$, $U$,
$B$, and $V$ filters on 27 December 2015.  The December observations were longer
integrations, excluded the $UM2$ filter, and had a non-detection in the shortest
wavelength filter, $UW2$.  %Full details are provided in Table~\ref{tab:swift}.
In both epochs, the $B$ and $V$ emission is spatially extended beyond
the PSF measured from nearby stars (see \S 3.1).   Other bands have too few
counts to detect any spatial extent.

The counts are measured from aperture photometry with $5^{\prime\prime}$
(radius) extraction regions and separately from Gaussian fits to the
object along the horizontal and vertical axes of the detector.
Aperture photometry produces 
results that are consistent with those obtained from
LCOGT, but with larger fractional
errors than Gaussian fits for marginal detections.  PSF photometry is also not applicable to
spatially extended emission.  In the Gaussian fits, counts are summed
over $3$ pixels in both directions
on the detector, are fit with one-dimensional (1D) Gaussian profiles, and then have
counts averaged between these two fits.
Results from both methods are listed in Table~\ref{tab:swift}.

\subsection{Ground-based Near-IR Photometry}
We observed ASASSN-15qi in the $JHK_s$ bands with the  NOTCam camera at
the Nordic Optical Telescope on 22 December 2015
(Table~\ref{tab:nirphot}).  The NOTCam data were reduced with 
an external IRAF package NOTCam version 2.5$^2$.
ASASSN-15qi was
also observed three times in the $H$ band with the Liverpool Telescope.
\footnotetext[2]{http://www.not.iac.es/instruments/notcam/guide/observe.html}

Bright, isolated, single stars  across the field  were used to
construct a characteristic PSF for each mosaic image.  The PSF
photometry of all the sources was obtained using the {\it allstar} task in
IRAF.  The zero-point magnitude was calculated using 2MASS photometry
of isolated stars and was then applied to ASASSN-15qi.  The 
calibration accuracy is $\sim 0.07$ mag for the NOTCam images, $\sim
0.1$ mag for the Liverpool Telescope images obtained on 28 December
2015 and 12 June 2016, and
$\sim 0.15$ for the Liverpool Telescope image obtained on 4 January 2016.

\subsection{Spitzer Mid-IR Photometry}

{\it Spitzer}/IRAC imaging of this region was obtained in all four bands
(3.6, 4.5, 5.8, and 8.0 $\mu$m) on 29 December 2006 (Program ID 30734,
PI Donald Figer).  We downloaded the corrected basic calibrated data (cBCD), 
processed the images with the automated pipeline, version S18.25.0,
and created mosaics with MOPEX  \citep[version 18.0.5,][]{makavoz2005}
at an image scale of $1\farcs2$ per pixel.

ASASSN-15qi is located where the ends of two filaments meet.
At 3.6 and 4.5 $\mu$m, the filaments do not contribute significantly
to the total flux.  The object flux was measured with point-response 
function (PRF) fitting using the tool Astronomical Point Source EXtraction  (APEX),
developed by the {\it Spitzer}  Science Center.  
However, at 5.8 and 8.0 $\mu$m, the object is heavily contaminated by
two filaments and also begins
to blend with a nearby object.  In these bands, both filaments are fit
separately with Gaussian
profiles along the horizontal axis in the mosaic.  The amplitude, centroid, and width of both
filaments are then calculated at the location of the source 
and are
subtracted from the image.  The source flux is then measured using 
point-source photometry, constrained by the location of ASASSN-15qi at
3.6 and 4.5 $\mu$m. This approach yields a detected flux
at 5.8 $\mu$m and an upper limit for a non-detection at 8.0 $\mu$m.
The final photometry is listed in Table~\ref{tab:nirphot}.

{\it WISE} W1 and W2 photometry of ASASSN-15qi is five times brighter than {\it
  Spitzer}/IRAC photometry at similar wavelengths.  The {\it WISE}
photometry of ASASSN-15qi is similar in three separate observations obtained over
500 days.  Since {\it WISE}
has a larger PSF than {\it Spitzer}, this emission is
likely contaminated by diffuse emission from the filament.
The {\it WISE} photometry is not included in our analysis.

\begin{figure*}[!t]
\plotone{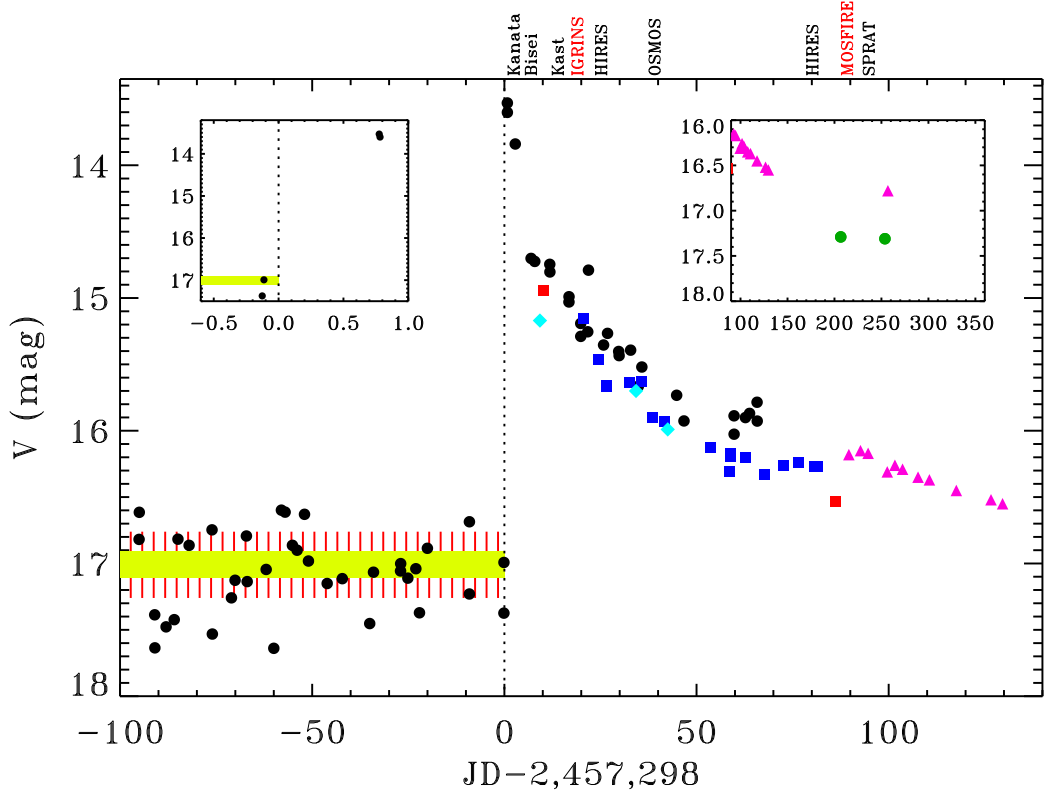}
\caption{The $V$-band lightcurve of ASASSN-15qi from ASAS-SN (black circles), LCOGT
(blue squares), {\it Swift} (red squares), Brimacombe (purple triangles), Liverpool
Telescope (green circles) and archival data (cyan).  The dotted line shows
the date that the outburst occurred.  The shaded yellow region shows
the average of the pre-outburst flux measurements and the instrinsic
variability of 0.1 mag.  Most of the scatter is caused by photon noise
near the sensitivity limit of 17 mag~for the ASAS-SN survey.  Dates
of spectroscopic observations are marked at the top of the plot.}
\label{fig:lc}
\end{figure*}
 
\subsection{JCMT/SCUBA2 Sub-mm Photometry}

We obtained James Clerk Maxwell Telescope
(JCMT) SCUBA2 sub-mm observations of ASASSN-15qi on 5 November
2015, 34 days after the outburst peak, as part of program M15BI083.
We also analysed archival 450 $\mu$m and 850 $\mu$m SCUBA-2
observations from project M11BGT01 obtained on 14 November 2011 and 
previously published by \citet{sreenilayam14}.  The
data were reduced using the standard Gould Belt Survey Legacy
Release 1 reduction parameters \citep{mairs15}.  The source was not
detected in either observation, with $3\sigma$ upper limits at 850
$\mu$m of 0.06 Jy/beam in the archival data and 0.07 Jy/beam in our
data (beam size of 14$^{\prime\prime}$).   At 450 $\mu$m
(8$^{\prime\prime}$ beam), the upper limits are 1.4 Jy/beam
and 3.4 Jy/beam.

\subsection{Low Resolution Optical Spectroscopy}

Low-resolution optical spectra were obtained at the 
Lick/Shane 3~m telescope with the Kast double spectrograph \citep{miller93},
at the MDM Telescope with
the OSMOS spectrograph, at the Liverpool Telescope with the SPRAT
Spectrograph, and with the Multi-Object Double Spectrograph at the
Large Binocular Telescope (LBT)   We also include in our analysis the low-resolution optical
spectra from the Kanata 1.5~m Telescope at Higashi-Hiroshima Observatory and the 1.01~m telescope at Bisei
Astronomical Observatory \citep{maehara15} and low-resolution near-IR
spectra from IRTF/SpeX \citep{connelley15}.  The Kast, OSMOS, 
SPRAT, and LBT spectra were obtained with the slit aligned with the parallactic
angle \citep{filippenko82}, while the Kanata and Bisei spectra were
obtained with the slit at a position angle of $0^\circ$.  The OSMOS,
SPRAT, LBT, Kanata, and Bisei spectra were all flux calibrated.

\subsection{Keck/HIRES Optical Spectroscopy}
Keck/HIRES \citep{vogt94} observations of ASASSN-15qi were obtained on 27 October 2015 and 21
December 2015.  The October
observations were obtained with the $0\farcs86\times7^{\prime\prime}$ slit
($R\approx45,000$) while the December observations utilized the
$1\farcs15\times7^{\prime\prime}$ slit ($R\approx36,000$), both chosen to match the
seeing of $0\farcs9$ and $1\farcs1$.  The spectra span 4800--9200 \AA\ with
some gaps between orders.  Spectra were reduced and calibrated with
the $makee$ pipeline written by T.~Barlow.
The region around the \ion{K}{1} $\lambda7699$ line was awkwardly corrected for telluric
absorption with a CFHT/ESPaDOnS spectrum of 72 Tau obtained on a
different night.

\subsection{Keck/MOSFIRE Near-IR Spectroscopy}
Keck/MOSFIRE \citep{McLean2012} long slit $K$-band spectra (2.0--2.3 $\mu$m) of ASASSN-15qi were obtained on 30
December 2015 at 1.6 airmasses.  The observations consist of four AB patterns with a total integration time
of 400 s.  The $12^{\prime\prime}\times1\farcs1$ slit was aligned at a position angle of
$4^\circ$.  The seeing of $0\farcs6$ led to a spectral resolution of
$\sim 3000$.  

The data were reduced with custom-written programs in IDL.  Telluric absorption was
corrected with a spectrum of the A0V star HIP 24311, which was observed at 1.4
airmasses.    The relative wavelength calibration from OH sky lines is
accurate to 3 \kms, although the absolute accuracy depends on object
centering in the slit.

\subsection{Harlan J. Smith 2.7~m/IGRINS near-IR Spectroscopy}

We obtained Target-of-Opportunity observations of ASASSN-15qi with
the Immersion Grating Infrared Spectrograph (IGRINS) on the
Harlan J. Smith Telescope at McDonald Observatory on 23 October 2015
(JD 2457318.598).  IGRINS is a high-resolution
near-IR echelle spectrograph providing $R\approx45,000$ spectra
simultaneously from 1.48--2.48 $\mu$m in 53 orders \citep{park14,gully12}.  ASASSN-15qi was
observed in two sets of $4\times10$ minute exposures in ABBA nodding
patterns with $7^\prime$ nods for a total exposure time of 80
minutes.  ASASSN-15qi was observed at $\sim 1.2$ airmasses, with an
A0V standard star observed at a similar airmass between the two ABBA nod patterns.

The $1^{\prime\prime}\times15^{\prime\prime}$  slit was fixed
to a position angle of 45$^\circ$.   The seeing was
$\sim 2^{\prime\prime}$ in the $H$ band and $1\farcs6$ in the $K$ band, as
measured in the spatial distribution of the continuum emission in the
cross-dispersion direction on the detector.  The spectral image was
rectified using Version 2.0 of the IGRINS pipeline$^3$.
Each pixel on the detector covers $0\farcs27$.  The spectrum was
extracted from a 9-pixel ($2\farcs4$) region centered on the star.
\footnotetext[3]{https://zenodo.org/record/18579\#.V7GE3T595YK}

\section{PROPERTIES OF ASASSN-15qi DURING QUIESCENCE AND OUTBURST}

Figure~\ref{fig:lc}  shows the $V$-band lightcurve of ASASSN-15qi, with
locations marked for each of our spectroscopic data points (see also
Tables~1--3).  
The ASAS-SN monitoring, which began on 
16 Dec.~2014, yields a quiescent (pre-outburst) average brightness of
$V=17.01$ mag.  The scatter in the pre-outburst photometry$^4$ is consistent with measurement
uncertainties combined with a variability of $\sim 0.05$ mag
($1\sigma$ standard deviation).
A day before the outburst,
two ASAS-SN telescopes measured $V=17.16\pm0.10$ mag.
\footnotetext[4]{The visual appearance of a larger scatter in the
  pre-outburst photometry is caused by the large statistical
  uncertainties in the individual measurement.}

On 2 October 2015, 23 hours after the previous observation,
ASASSN-15qi was found to have brightened to $V=13.53$ mag, an increase of a factor of $\sim 25$ in
flux.  
The brightness increase was at least $\sim 3.5$ mag and may have been
larger because the ASAS-SN photometry includes all nebulosity within
$15^{\prime\prime}$.  The inclusion of nebulosity would increase the pre-outburst
brightness measurements much more than the brightness near the
outburst peak.  
 Indeed, the latest photometric data from the Liverpool
Telescope, obtained 254 days past outburst (11 June 2016), 
has $V=17.31$ mag when measured in a $5^{\prime\prime}$
radius aperture and $V=17.52$ when measured with PSF photometry.

\begin{figure*}[!t]
\epsscale{0.75}
\plotone{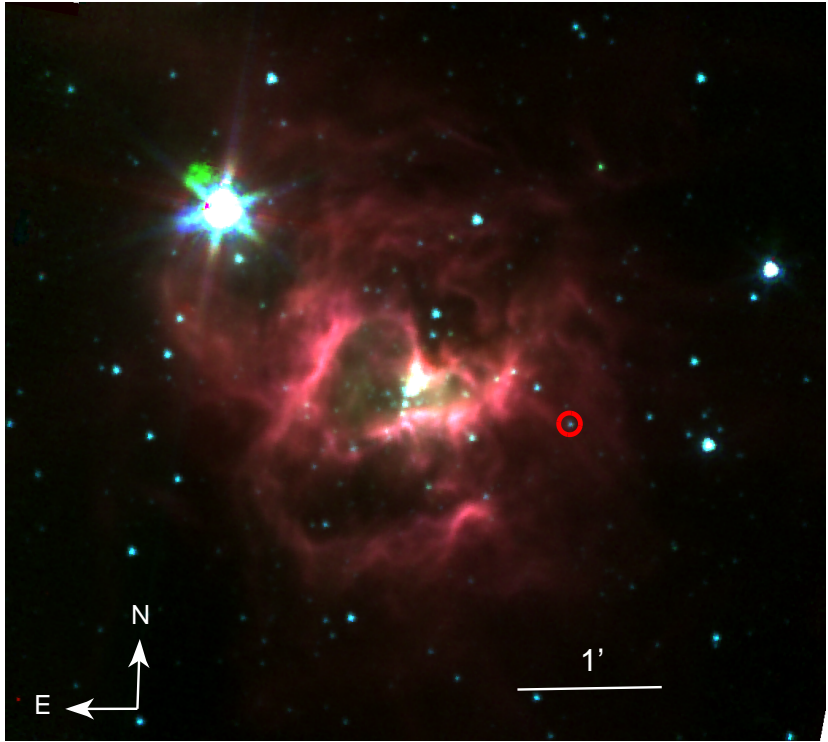}
\caption{{\it Spitzer}/IRAC 3.6/4.5/5.8 $\mu$m image, showing the location of ASASSN-15qi (circle) near the star formation complex Sh2-148.}
\label{fig:spitzerimage}
\end{figure*}
 
\begin{figure}[!h]
\epsscale{1.}
\plotone{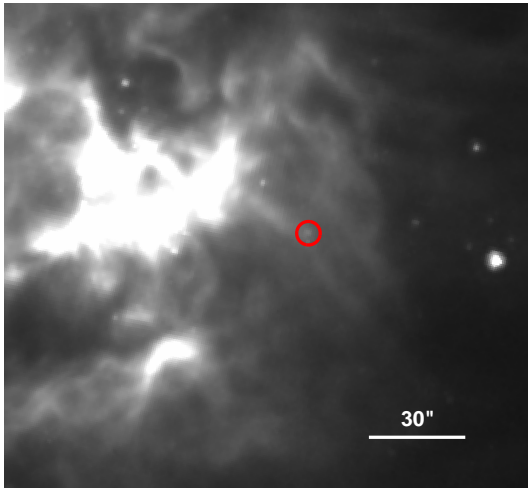}
\caption{{\it Spitzer}/IRAC 5.8 $\mu$m image of the immediate
  region around ASASSN-15qi
  (red circle).  The object is located at the location of a projected
  bend in a warm dust filament.}
\label{fig:filament}
\end{figure}
 
The flux then decayed quickly.  Fits to the flux lightcurve show an initial $e$-folding decay time from peak of
$\sim 6$ days, based on few data points and assuming that the first
post-outburst photometry is during this decay and not during the
initial rise.  After this initial
rapid decay, ASASSN-15qi faded with an $e$-folding time of $\sim
50$ days.  The object then settled at $V=16.2$ mag for $\sim 30$ days
before resuming its decay, with an $e$-folding time of $\sim 80$ days.
The flattening, delayed continuation of
the decay, and the possible bump at 80--100 days may all be related to
the light travel time to the nebulosity around the object.
Differences in the ASAS-SN and LCOGT photometry between days 50--70
are likely related to the inclusion of nebulosity in the large ASAS-SN
pixels and extraction aperture.  The photometric measurements from
observations with small pixels and better seeing should provide a more direct measure of the
outbursting source itself (see also photometry with different
extraction methods listed in Table~\ref{tab:photometry}.

In the following subsections we describe our analysis of
this outburst using this lightcurve and the photometric and
spectroscopic follow-up observations described in \S 2.  These sections frequently describe the days
past outburst peak, defined as JD 2,457,298.  The object is described as
in quiescence during the pre-outburst phase.

\vspace{7mm}

\subsection{Location in the Molecular Cloud}

\begin{figure}[!t]
\vspace{-4mm}
\epsscale{1.}
\plotone{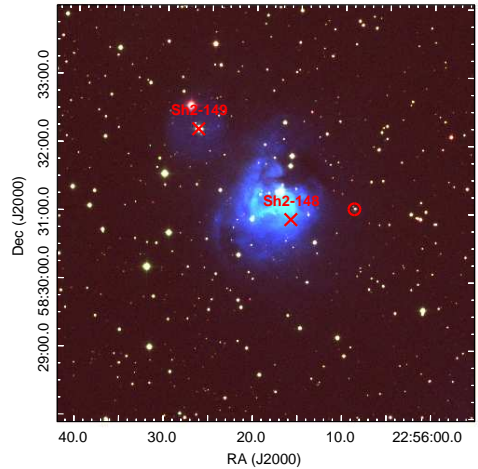}
\caption{IPHAS $r$-band image (with H$\alpha$ emission in blue)
  shows ASASSN-15qi near the central positions of the \ion{H}{2}
  regions Sh 2-148 and Sh 2-149 (marked locations measured by Azimlu
  \& Fich~2011).}
\label{fig:iphas}
\end{figure}

Figures ~\ref{fig:spitzerimage}--\ref{fig:iphas} show {\it Spitzer}/IRAC infrared and IPHAS optical
images of the
region around ASASSN-15qi.  Bright H$\alpha$ nebulosity, warm dust
emission, and bright 850 $\mu$m emission are all centered around the
position of Sh 2-148, with a central location $1^\prime$ from ASASSN-15qi.  
As projected on the sky, the object
is located along a filament of excess dust emission, 
either at a bend in a single filament or at the projected location where
the ends of two filaments meet.
A visual companion located $2\farcs7$ NE is 2.8 mag fainter than the
quiescent optical brightness of
ASASSN-15qi, as measured in the IPHAS archive \citep{barentsen14}, and likely does not contribute to any
of the observed variability.  In optical images, some nebulosity is detected just south of the
target, as described in \S 3.1.1.

The
cloud velocity of $v_{\rm helio}=-65$ \kms ($v_{\rm lsr}=-54$ \kms) has been
measured with sub-mm CO emission from the nearby \ion{H}{2} regions
\citep{azimlu11}, and is adopted as the stellar radial velocity.$^5$  This
velocity is consistent with the velocity of emission and absorption
lines from the nearby nebulosity (\S 3.1.2) and with the velocity of
photospheric lines (\S 3.4).  The cloud velocity is used instead of
the stellar velocity because the
detected photospheric lines are very broad, have low signal-to-noise ratios, and may
suffer from velocity offsets introduced by the wind.
\footnotetext[5]{All interstellar velocities listed
below are in the heliocentric frame of reference.  Discussions of wind and emission
components use velocities relative to the adopted stellar radial
velocity of $-65$~\kms.}

\subsubsection{Nebulosity in Optical Imaging}

Initial follow-up imaging did not show any
nebulosity nine days after the outburst peak \citep{stecklum15a}.  However, some extended emission was
detected to the SE of the outburst in slit acquisition images for
the Keck/HIRES spectra obtained 26 days
after the outburst peak by \citet{hillenbrand15}.

\begin{figure*}[!t]
\epsscale{1.}
\plottwo{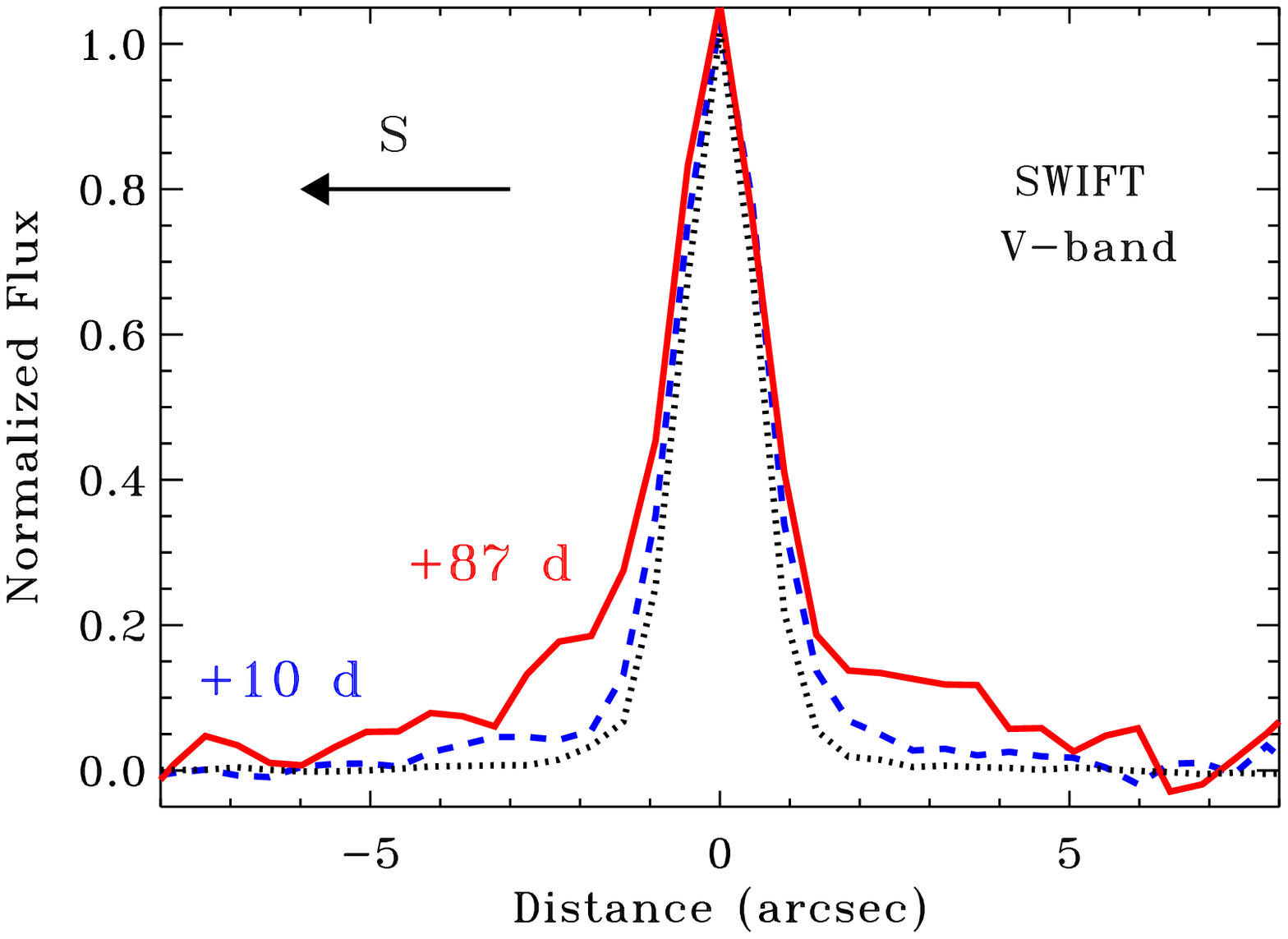}{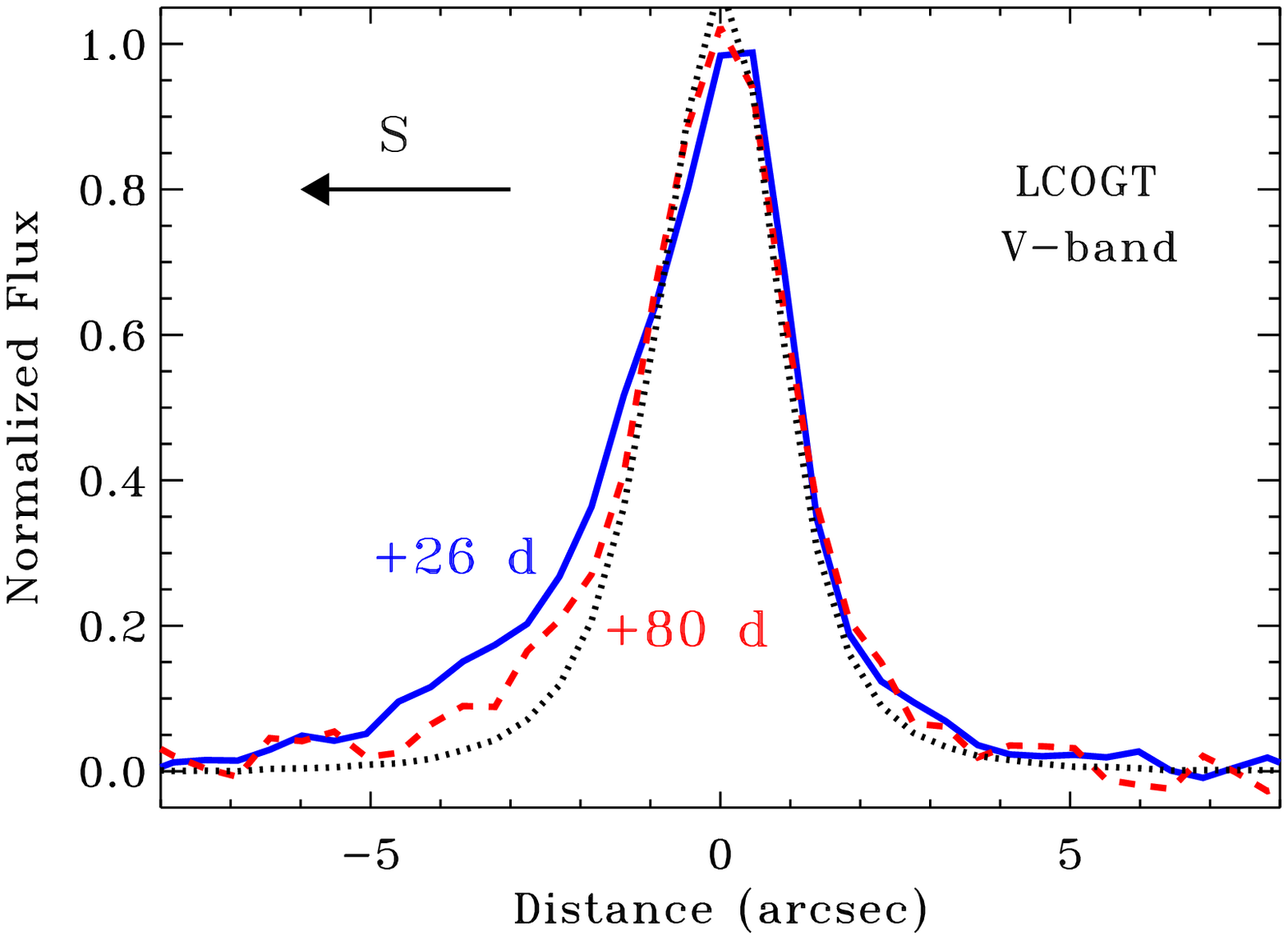}
\caption{The spatial profile in the N-S direction of $V$-band emission observed with {\it Swift}
  (left, 10 and 87 days after outburst peak, $V=14.94$ and $16.53$, respectively) and LCOGT (right, 26 days and 80 days after outburst peak,
  $V=15.66$ and 16.27, respectively) on different dates.  
The emission from ASASSN-15qi is consistent with the PSF (dotted
lines) in the {\it Swift} image obtained 10 days after the outburst peak.
The emission is then extended beyond the PSF in images obtained on
subsequent dates.  After 80 days, the spatially
extended emission is fading, along with emission from the central star.
 The seeing was $2\farcs2$ for both LCOGT
observations shown here.
}
\label{fig:nebemission}
\end{figure*}

The nebulosity also appears in our LCOGT and {\it Swift} imaging with a
time dependence consistent with both the \citet{stecklum15a}
non-detection and the \citet{hillenbrand15} detection.
Figure~\ref{fig:nebemission} shows $V$-band nebulosity in LCOGT and {\it Swift}
images as a 1D spatial distribution, extracting counts from S to N
along the direction of the nebulosity$^3$.  The first {\it Swift} $V$-band image,
obtained 10 days after outburst peak, includes little contribution from
nebulosity, although this emission is detected by 87 days after peak.   LCOGT images obtained 26 days after outburst
show strong nebulosity, which then weakened in later epochs of LCOGT
imaging.
By day 80 after the outburst peak, the nebulosity faded
relative to the central object, and the central object itself had
faded by 0.6 mag since day 26.  This comparison is qualitatively consistent with
results on other dates and bands.

The nebulosity is likely a reflection nebula or a variable \ion{H}{2} region that brightened with
the central source.  The delay between outburst and nebulosity
brightening and the flattening in the lightcurve (see
Figure~\ref{fig:lc}) are likely caused by light travel time.  Given
the estimated distance, light from the
star would take 35 days to arrive at a projected distance of $\sim 2^{\prime\prime}$
(the approximate centroid location of the nebulosity projected
on the sky).  A more complete analysis of the nebulosity brightness versus time
is beyond the scope of this
paper.

\subsubsection{Nebulosity in Emission Lines}

H$_2$ emission is detected in all near-IR spectra.  While the H$_2$
1-0 S(1) line is most prominent, a rich forest of other H$_2$ lines
is also present in the high resolution IGRINS spectrum (see Appendix
A for a line list and description of the fits).

\begin{figure}[!t]
\epsscale{1.}
\plotone{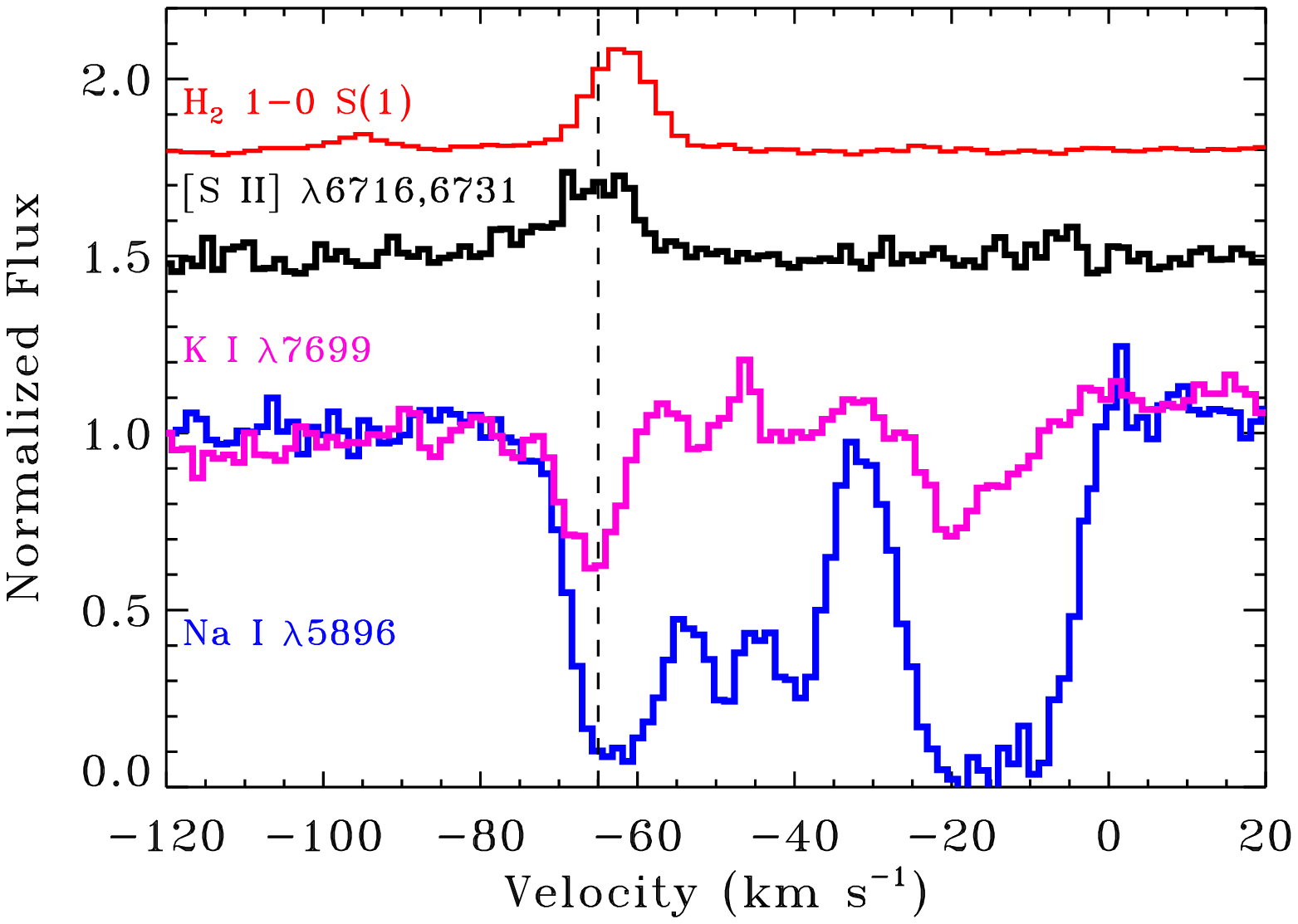}
\epsscale{1.}
\plotone{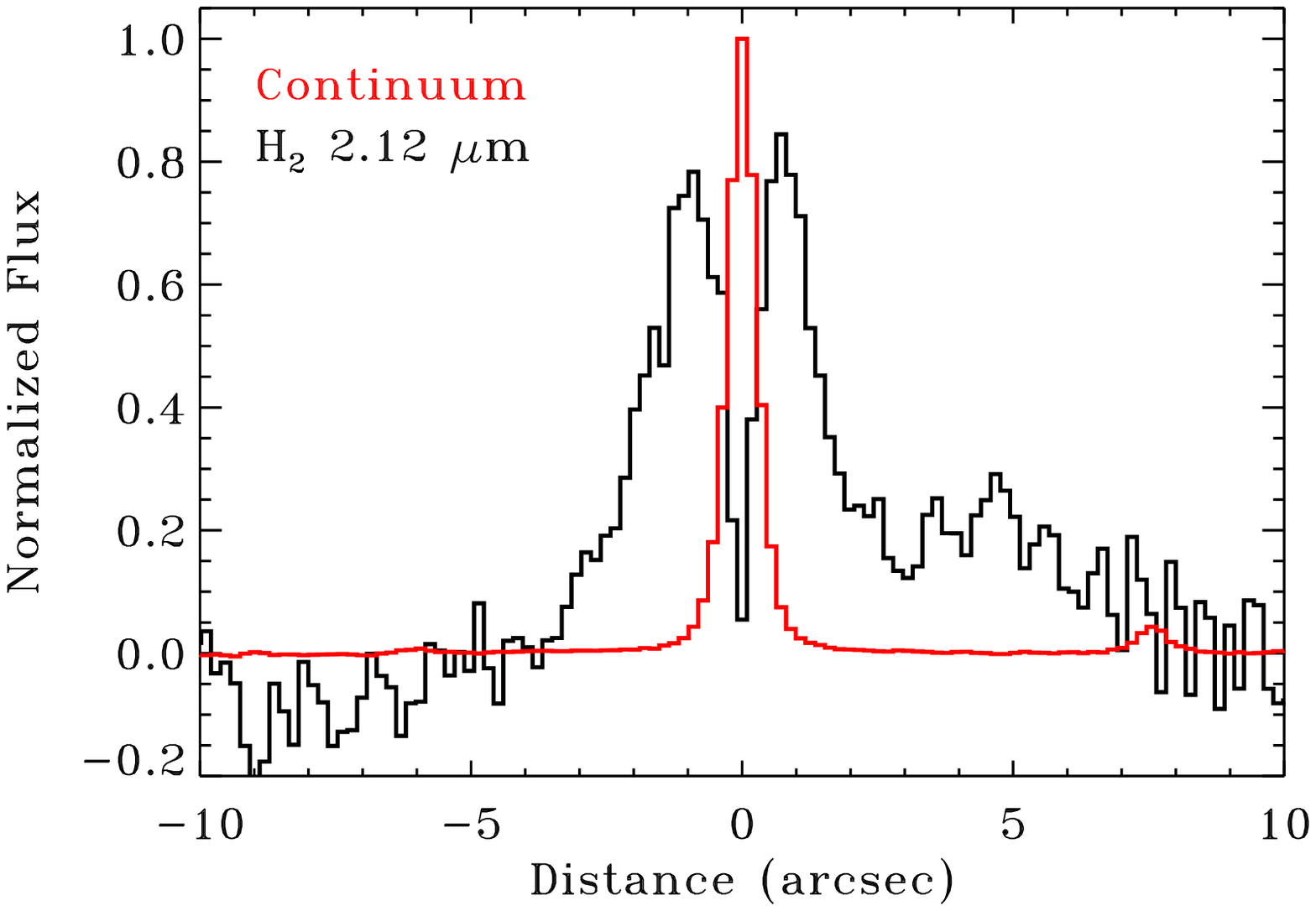}
\caption{Top:  The molecular cloud and nebulosity seen in H$_2$ and
  [\ion{S}{2}] in emission and \ion{K}{1} and \ion{Na}{1} in
  absorption. Bottom:  The spatial distribution of H$_2$ 1-0 S(1) and 2.12 $\mu$m
  continuum emission along the slit in the Keck/MOSFIRE spectrum.}
\label{fig:h2vel}
\end{figure}

The average
H$_2$ line centroid velocity of $-62$ \kms\ is consistent with the
heliocentric radial
velocity of the parent molecular cloud (left panel of Figure~\ref{fig:h2vel}).
Emission is detected in transitions from the common 1-0 S(1)
line at 2.1218 $\mu$m to lines with high vibrational excitation,
including 8--6 O(4) and 9--7 Q(2).  Excitation to such high
vibrational levels requires a strong ultraviolet (UV) field.  Many
of the detected emission lines, such as 8--6 O(4), are prominent
indicators of UV fluorescence, as predicted in the models of \citet{black87}.
  Excitation to similarly high vibrational levels has previously been
observed in the \ion{H}{2} region around HD~37903, a B1 star in Orion
\citep{burton98,meyer01}, and is also found in spectra of planetary
nebulae \citep[e.g.,][]{hora99}.   H$_2$ emission from lower vibrational levels is
common in FUor objects \citep[e.g.,][]{aspin11}.

The H$_2$ emission is spatially resolved in the cross-dispersion
direction in both the IGRINS and MOSFIRE
spectra (Figure~\ref{fig:h2vel}).  In the MOSFIRE spectrum, which was obtained in better
seeing, H$_2$ emission is not detected from the star, has two peaks
at $\pm0\farcs5$ off the star, and becomes undetectable at
$4^{\prime\prime}$ (0.1 pc) S and $7^{\prime\prime}$ 
(0.2 pc) N.  The Br$\gamma$ and CO overtone emission is not spatially
extended beyond the continuum emission.  The
H$_2$ lines are extended by $\sim 2^{\prime\prime}$ beyond the
continuum emission.  In the IGRINS spectrum, the
H$_2$ 1-0 S(1) centroid is located $0\farcs25$ from the continuum
emission in the SW direction along the slit, while other H$_2$ lines
are offset by $0\farcs8$ from the continuum in the same direction.
The weaker, low excitation lines are also located $\sim 2$ \kms\
redward of H$_2$ 1-0 S(1) and other low excitation lines.  Both the
velocity and spatial information suggest multiple components, one of
which has strong UV excitation.

Emission is only marginally detected in the H$_2$ 1-0 S(1) line in the SpeX
spectrum of \citet{connelley15}, which was obtained 13 days past
outburst peak.  The equivalent width (EW) of $0.06\pm0.03$ \AA\ is a factor of $\sim 20$ weaker
than the EWs of 1.2 and 0.81 \AA\ in
the subsequent IGRINS and MOSFIRE spectra, respectively, despite
little change in the $K$-band continuum flux. Comparisons between these spectra
are challenging because of different seeing conditions and slit
widths.  However, the SpeX image was obtained with seeing of
$0\farcs8$ and a $0\farcs8$ slit width, so the results should be
similar to that of the MOSFIRE spectrum.  The H$_2$ 1-0 S(1) emission is likely variable,
with a time delay similar to that seen in the brightness of the
optical nebulosity.

Emission is also detected in the forbidden [\ion{N}{2}]
$\lambda6583$ line and [\ion{S}{2}] $\lambda\lambda$6716, 6731 doublets.  The
lines are located at $-66$ \kms.  The [\ion{S}{2}] equivalent widths increased by a
factor of $\sim 2.9$ from 26 days past the outburst peak to 80 days
after the peak, when the object was near quiescence.  These equivalent
widths 
are approximately consistent with a constant line flux against a weaker
continuum.  The [\ion{N}{2}] line is stronger during outburst than
during the decay.  Emission is not detected in either [\ion{O}{1}] or [\ion{O}{3}] forbidden lines.

A number of atomic lines were detected in the SpeX near-IR spectrum of
\citet{connelley15}.  These lines are generally too weak in the IGRINS
spectrum for a robust line identification and analysis.

\begin{figure*}[!t]
\plotone{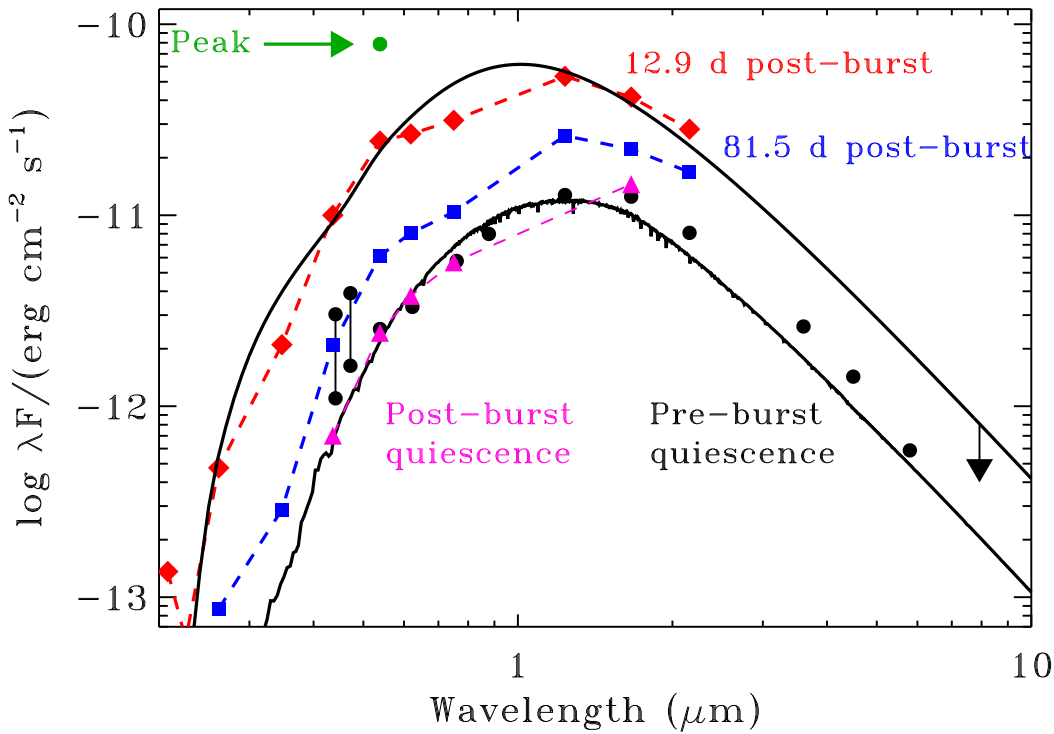}
\caption{The SED of ASASSN-15qi during
  quiescence (black circles) and near the outburst peak (red squares),
  during the decay (blue squares), and in the final photometry once
  the source faded to quiescence (magenta triangles).  The outburst peak is shown
  only in the $V$ band.  The quiescent SED is
  well fit at red/near-IR wavelenths by a 6200 K photosphere with
  excess emission at blue wavelengths.
 The outburst spectrum is
  reasonably well fit by  10,000 K blackbody (black lines).  Both fits
assume an extinction of $A_V=3.6$ mag, as measured in \S 3.2.3.}
\label{fig:sed}
\end{figure*}

\begin{figure}[!hb*]
\hspace{-5mm}
\epsscale{1.}
\plotone{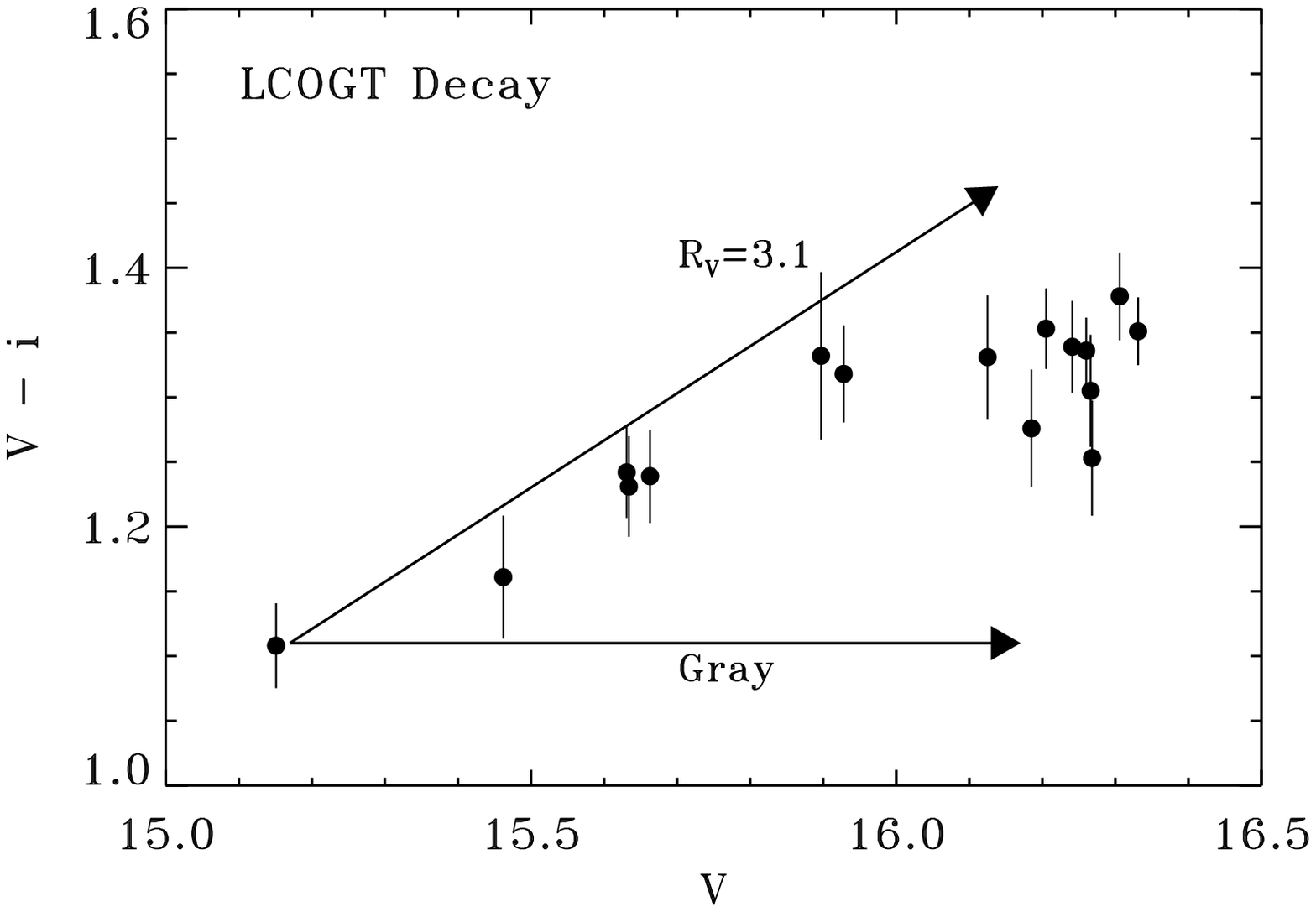}
\plotone{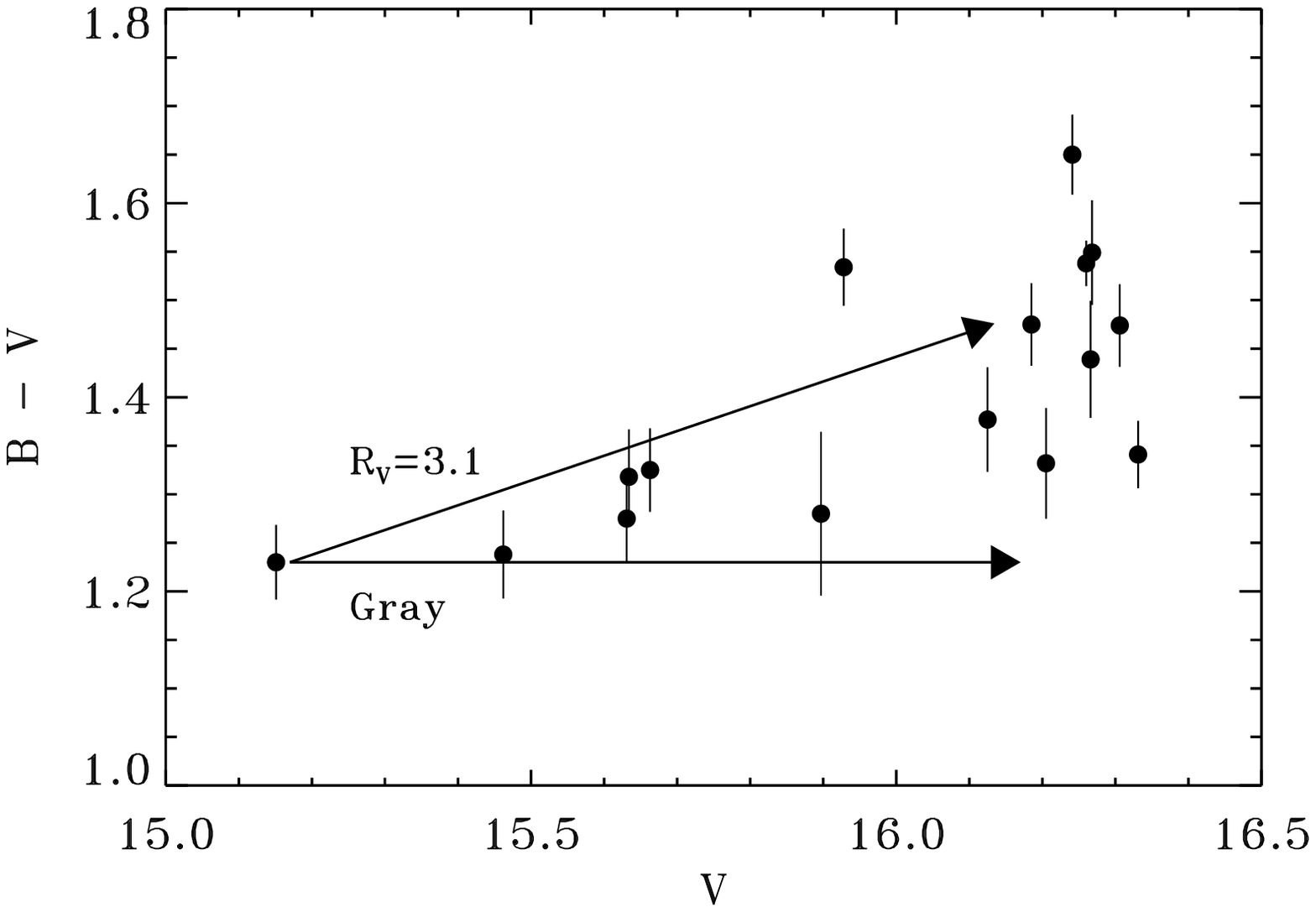}
\caption{The $V-i$ (top) and $B-V$ (bottom) color of the outburst versus the brightness,
  with reddening vectors for standard interstellar medium extinction and gray extinction.  The $V-i$ color becomes
much redder during the decay while the $B-V$ color remains roughly
constant.  The errors are statistical uncertainties that affect relative
colors and do not include the systematic
errors that would affect the absolute scaling of the colors.}
\label{fig:color}
\end{figure}
 
\subsubsection{The Interstellar Environment in Absorption}

Interstellar absorption lines are detected in the
\ion{Na}{1} $\lambda\lambda$5889, 5996 and \ion{K}{1} $\lambda7699$ lines$^6$
(see Figure~\ref{fig:h2vel}),
tracing either the nearby nebulosity or the parent molecular cloud.
The \ion{Na}{1} lines are more optically thick and include more
absorption components than the \ion{K}{1} line.  The \ion{K}{1} line has narrow absorption components
centered at $-66$ \kms, likely local to the star, and at $-18$ \kms\ in a
cloud closer to us.
\footnotetext[6]{The \ion{K}{1} $\lambda7665$ line is not located on the HIRES
  detector in the utilized spectrograph configuration, and in any case
  is located within a mess of telluric absorption lines.}

%LIC:  -24.5 KM/S
%G-cloud:  -27.4 km/s

Diffuse interstellar bands (DIBs) are also detected at several wavelengths,
including 5780, 5797, 6177, 6283, and 6614 \AA, with velocities
roughly consistent with the parent molecular cloud.  The equivalent width
of the 5780 \AA\ DIB is consistent with $A_V=2.2-3.5$ mag
\citep{friedman11}, with the caveat that their sample does not include
lines-of-sight that are higher than $A_V=3.5$ mag.

The depth and velocity of the absorption components attributed to the
interstellar medium remained
constant in both lines and diffuse interstellar bands between our first HIRES
observation, obtained 25 days after the outburst peak ($V\approx15.55$ mag)
and our second HIRES observation, obtained 80 days after
the oubtburst peak ($V\approx 16.4$ mag).

\begin{table}
\caption{Outburst SED$^a$}
\label{tab:phot}
\begin{tabular}{cccccc}
\hline
& Pre-burst & \multicolumn{3}{c}{Outburst: days past peak} & Post-burst\\
Band & Quiescence & 0 d & 12.9 d & 81.5 d & 254 d\\
\hline
$UVW2$ & --  & -- & 19.1  & --\\
$UVM2$ & --  & -- & 20.7  & --\\
$UVW1$ & --  & -- & 18.0  & 20.2\\
$U$       & --  & -- & 16.8   &  18.8\\
$B$ & 17.4--18.2 & --  & 16.0     & 17.7 & 19.0\\
$g$ & 17.06--18.01 & -- & --\\
$V$ & 17.01   & 13.5 & 14.8    & 16.3  & 17.3\\
%R & --   & --   & --   &  --\\
$r$ & 16.66 & --   & 14.4   & 15.7 & 16.5\\
$i$ & 15.59 &--    & 13.8   &  15.0 & 15.7\\
$I$ & 15.08 &    --&   --    &  \\
$J$ &  13.704 & --& 12.15 & 12.93 & 13.5$^b$\\
$H$ & 12.923 & --& 11.62 & 12.30 & 12.76\\ 
$K$ & 12.647 & -- & 11.29 & 11.85 & 12.2$^b$\\
$I1$ & 12.36  &--&-- &-- \\
$I2$ &  12.30  &--&--&-- \\
$I3$ &  12.50    &--&--&-- \\
$I4$ &  $>11.9$  &--&--&-- \\
\hline
\multicolumn{6}{l}{$^a$All numbers are mag in their native systems.}\\
\multicolumn{6}{l}{$^b$Estimated from $H$-band and near-IR colors.}\\
\end{tabular}
\end{table}

\subsection{The SED and Photospheric Emission}

In this section, we describe the emission from the optical
photosphere using spectra and the spectral energy distribution (SED).  The section is divided
into pre-outburst quiescence, the outburst and decay (0-150 days post-burst), and the final
epoch (about 250 days post-burst).  Table~\ref{tab:phot} presents the
SEDs in each epoch, while spectra and analysis are shown in Figures~\ref{fig:sed}-\ref{fig:balmerabs}.

\subsubsection{The SED before the 2015 outburst}

The pre-outburst SED of ASASSN-15qi is constructed from the {\it Spitzer}, 2MASS, IPHAS, {\it HST}
Guide Star Catalogue, Palomar Transient Factory, and pre-outburst ASASSN-15qi
photometry.  The archival pre-outburst photometry at blue wavelengths varies within a range
of $\sim 1$ mag, including $g$-band emission obtained over 4.5 months by the
Palomar Transient Factory.  Whether this level of variability is also
seen at redder wavelengths is uncertain.

The quiescent SED from 0.6--4.5 $\mu$m is reasonably fit with a synthetic
spectrum of 6200 K \citep{allard14} and an extinction $A_V\approx3.6$
mag, using an extinction curve from Cardelli et al.~1989 and
total-to-selective extinction value of $R_V=3.1$.  The effective
temperature and extinction are adopted from a measurement of the
post-burst spectrum (see \S 3.2.3). 
These values, the 2MASS $J$-band magnitude, and the
distance estimate of 3.24 kpc lead to a
quiescent luminosity of 18 L$_\odot$ for this component.  The
luminosity is consistent with the expectation that ASASSN-15qi is
significantly fainter than the nearby O9/B0 stars that power the compact
\ion{H}{2} regions Sh 2-148 and Sh 2-149 \citep{crampton78}, but still
bright enough and hot enough to excite the nearby molecular cloud material.
%No excess emission is detected in the mid-IR IRAC bands.

Excess emission is detected in $B$ and $V$ during quiescence, pointing
to the presence of hot emission in excess of a single temperature
fit.   The quiescent SED may alternately be reproduced with hot ($\sim
15,000$ K) and cool ($\sim 5000$ K) components.  The hot component
would need to be fainter than the cool component, either because of a
higher extinction or because of a smaller surface area.  The hot
component could have a smaller surface area if a hot star is
surrounded by a warm gaseous disk, if the disk has a small hot spot,
if the star has an accretion-induced hot spot, or in an evolved binary
system consisting of a white dwarf with a dwarf or giant companion.

The quiescent SED provides a baseline for interpreting past
photometry.  As discussed above, the pre-outburst ASAS-SN photometry
indicates a stable $V$-band magnitude.  
Photometry in many previous epochs is also consistent with a
quiescent SED, including epochs in 1953.83 and 1989.67.  However,
$g$ and $B$ photometry shows variability of $\sim 1$ mag.  In
1976.5, USNO-B recorded photometry that was $\sim 2$ mag brighter than
the quiescent SED, indicating a likely outburst.

\subsubsection{Photospheric emission during the outburst and decay}

The UV-optical-IR SED during the outburst is constructed at 13 and 81
days past the outburst peak.  These dates are selected to coincide
with the dates when $JHK$ photometry was obtained.  The multi-band LCOGT
lightcurve (Figure~\ref{fig:color}) is then used to estimate the $BVri$
magnitudes.  The UV photometry from {\it Swift} was obtained
contemporaneous to the $JHK$ photometry, and are listed with only minor adjustments to
account for temporal changes.

The LCOGT photometry demonstrates that the object was bluer
during the outburst than in quiescence (Figure~\ref{fig:color}).  The $B-V$ and $V-r$ colors
are roughly constant during the outburst decay and become redder when
the decay flattens out.  The $V-i$ color becomes much redder as the
outburst decays and flattened when the lightcurve flattened.
The few epochs with
$JHK_s$ photometry have bluer near-IR colors during outburst than during
quiescence.

\begin{figure*}[!t]
\epsscale{1.02}
\plotone{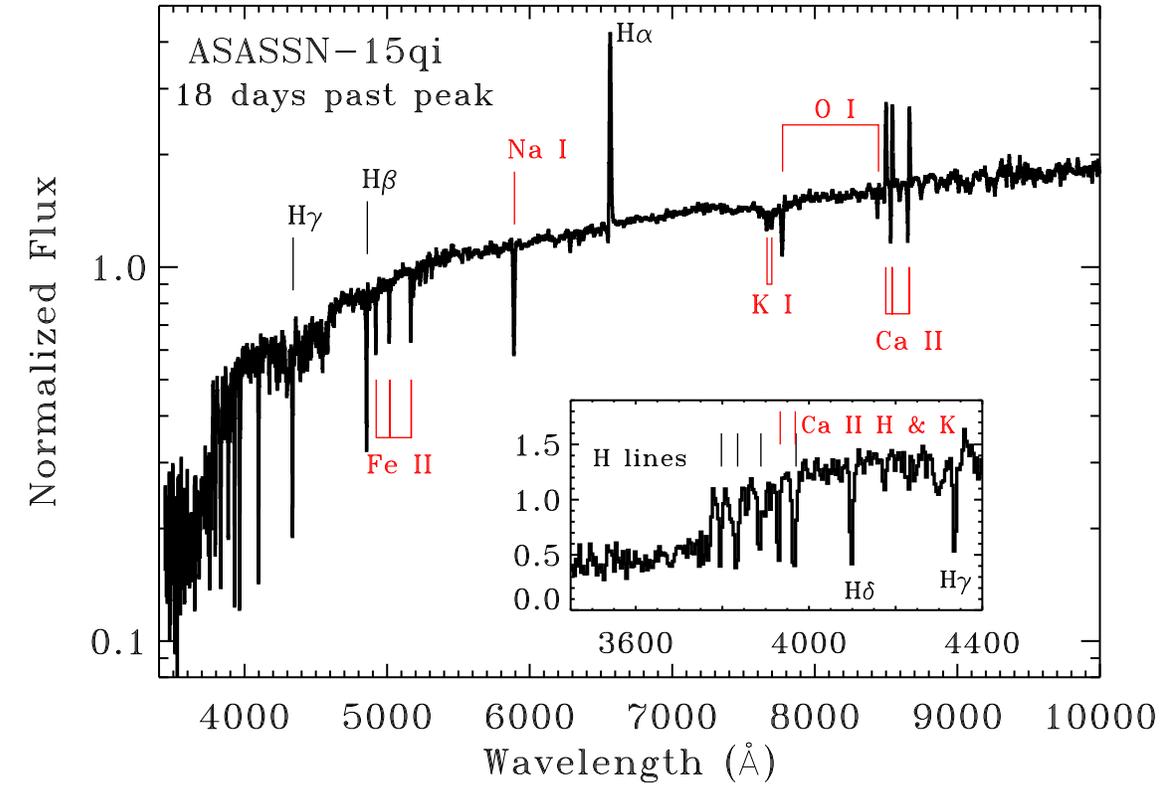}
\plotone{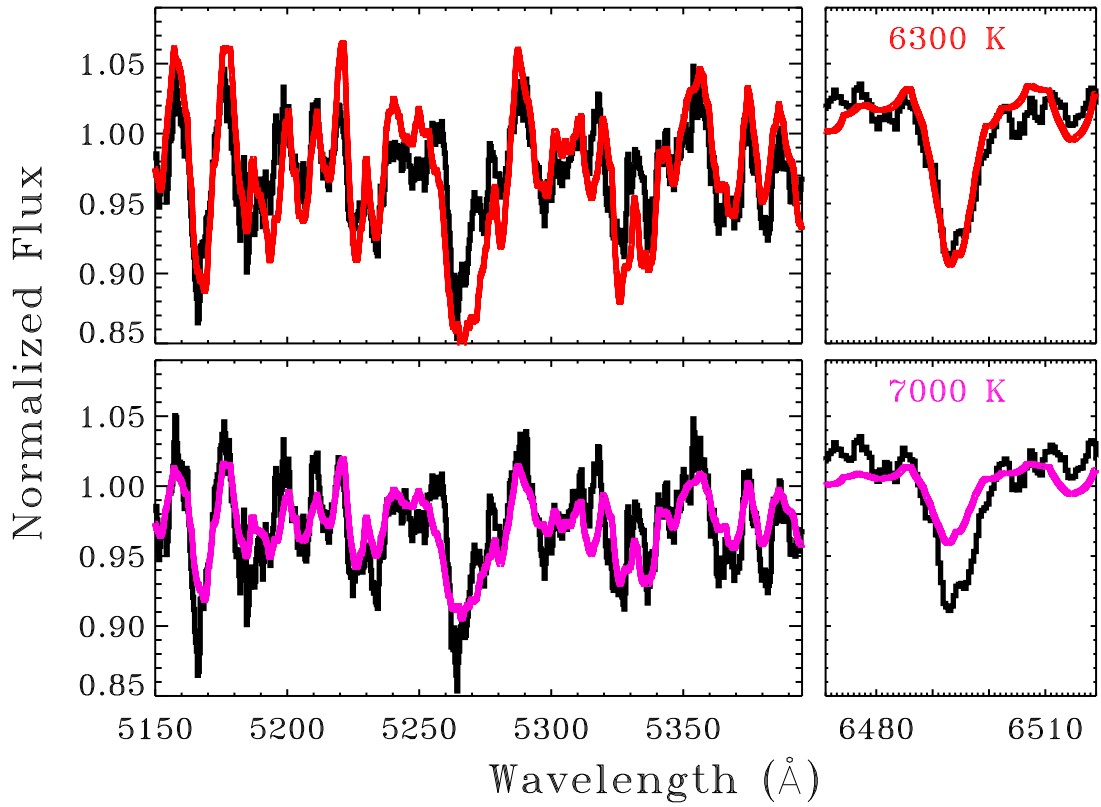}
\caption{Top:  The Lick Shane/Kast spectrum of ASASSN-15qi, with prominent
  H Balmer and metal lines marked in black and red, respectively.
  Bottom:  The binned Keck/HIRES spectra obtained in December during the decay (79 days
  post-outburst) compared to the BT-Settl spectral models of \citet{allard14}.  The
  5300 \AA\ region is well fit by a 7000 K photosphere, while the
  region around the  blend at $6497$ \AA\ is well fit by a 6300 K photpshere.}
\label{fig:speclo}
\end{figure*}

The SED at 13 days after the outburst peak is reasonably well fit with
a $10,000$ K blackbody of $320$ L$_\odot$ that suffers from an
extinction $A_V=3.6$, as shown in Figure~\ref{fig:sed}.  Adjusting this
luminosity to the peak photometry would lead to $L\approx1000$
L$_\odot$.  This luminosity increases is larger than the brightness
increase in the $V$-band because of the different bolometric
corrections in quiescence and outburst.  Alternately, if we assume
that the temperature does not change, then the luminosity
could be as low as 500 L$_\odot$ at peak and $180$ L$_\odot$ at 13
days post outburst -- although this scenario leads to a poor fit to
the SED.  If the brightness change were interpreted as attributed to
extinction, then the SED could be fit with $6200$ K emission and
$A_V=1.5$, with a luminosity only three times higher than
quiescence.  However, the extinction scenario is unlikely (see \S 4.2.2).
These scenarios all overestimate
the $i$-band photometry during the outburst.

The high resolution optical spectra obtained during outburst lead to
photospheric temperatures that are mostly consistent with the 6200 K temperature measured during
quiescence  (Figure~\ref{fig:speclo}).  The \ion{Fe}{1} and \ion{Mg}{1} line complexes at
5100--5400~\AA\ and the \ion{Fe}{1}+\ion{Ba}{2} blend at 6497~\AA\ are the
most identifiable photospheric features.  We fit these features with a
photospheric spectrum produced by BT-Settl models with solar
metallicity \citep{allard14}.  The 6497~\AA\ feature is
well reproduced by a 6300~K photosphere with $v \sin i=180$ \kms\
centered at the expected radial velocity ($-65$ \kms\ in the
heliocentric frame).  The observed spectrum near
5200~\AA\ is  well reproduced by a low gravity 7000 K photosphere with
a similar velocity broadening and centroid.

This 6000-7000 K photospheric temperature seen in both HIRES spectra
obtained during outburst poses a
challenge to the SED models, where the $V$-band emission is expected
to be produced by the hotter ($\sim 15,000$ K) component.  Morever, the single temperature fits
fail to reproduce some of the SED.  Most likely,
the assumption of a single temperature is overly simplistic.
Especially if the emission is produced in a gaseous disk with a
temperature gradient, the emission may instead arise from a range of
temperatures.  
As confirmation, the Balmer continuum absorption in the low resolution 
Lick Shane/Kast spectrum is consistent with a B5--B9 star when compared to the
\citet{pickles98} spectral library --- which suggests temperatures of
$10,000-15,000$ K, much hotter than the $\sim 6500$
K obtained from the high resolution spectra of the photosphere.

\begin{figure*}
\vspace{-146mm}
\epsscale{1.1}
\plotone{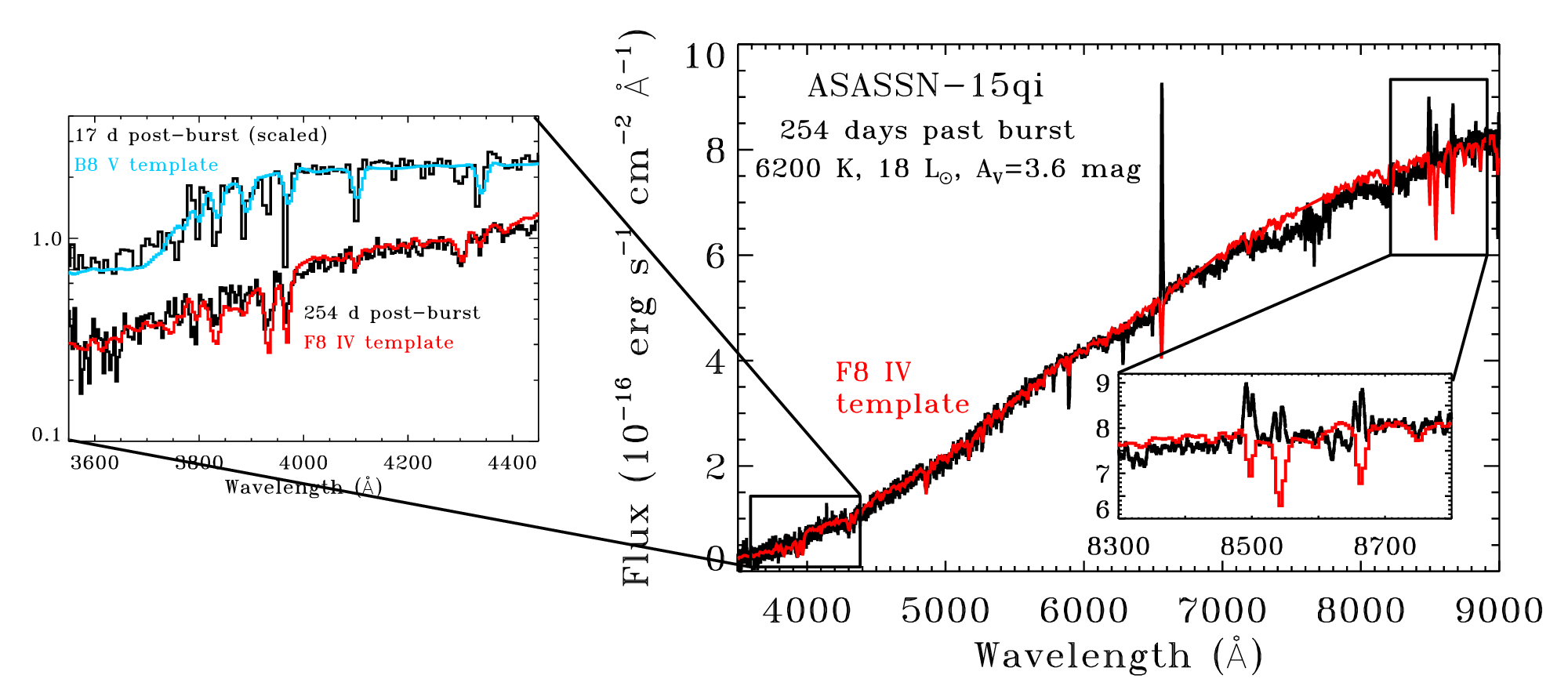}
\caption{The LBT/MODS spectrum of ASASSN-15qi obtained after the outburst
  had faded.  The insets zoom in on the \ion{Ca}{2} infrared triplet
  region (main plot) and the Balmer continuum region (left).  The quiescent spectrum is consistent with an F8 IV
  template \citep{pickles98} and $A_V=3.6$ mag (shown in red in all
  panels).  The Kast spectrum obtained 18 days after the outburst peak
  (left, top black spectrum) is consistent with a B8 V template.  The
  Balmer continuum absorption disappeared after the burst had decayed.}
\label{fig:balmerabs}
\end{figure*}

\subsubsection{Post-outburst quiescence}

In two late epochs with photometry obtained at 207 and 254 days past
outburst, the source brightness was constant.  By these dates, the outburst
had subsided and most signatures of the burst had disappeared.  In
this epoch, the
SED had a similar shape but is fainter at blue wavelengths than the SED inferred
from pre-outburst photometry.

The LBT optical spectrum obtained in this epoch is consistent
with an F8 star (Figure \ref{fig:balmerabs}), though some residual features of the wind that will be
discussed in \S 3.4.  The extinction of $A_V\approx 3.6$
mag measured from the spectrum is similar to $A_V\approx 4.0\pm0.2$ measured from the $V-H$
color.  The extinction $A_V\approx 3.6$ is adopted throughout the
paper.  This measurement may be affected by wavelength dependence in
the scattered light within the extraction aperture.   The total
luminosity of the object is $18$ L$_\odot$, as measured from fitting a
template 6200 K photosphere to the observed spectrum.

The outburst SED and the Balmer continuum absorption in the Kast
spectrum both demonstrate the presence of hot gas during the outburst.
The late-time spectrum no longer exhibits Balmer continuum
absorption (see inset in Figure \ref{fig:balmerabs}).  The $B-V$ color is also redder than the pre-outburst and
outburst photometry.  The hot ($\sim 10,000-15,000$ K) component that
was present during outburst, and that may have contributed to
short-wavelength emission before the outburst, is no longer visible in
the system.

\begin{figure}[!t]
\hspace{-8mm}
\epsscale{1.15}
\plotone{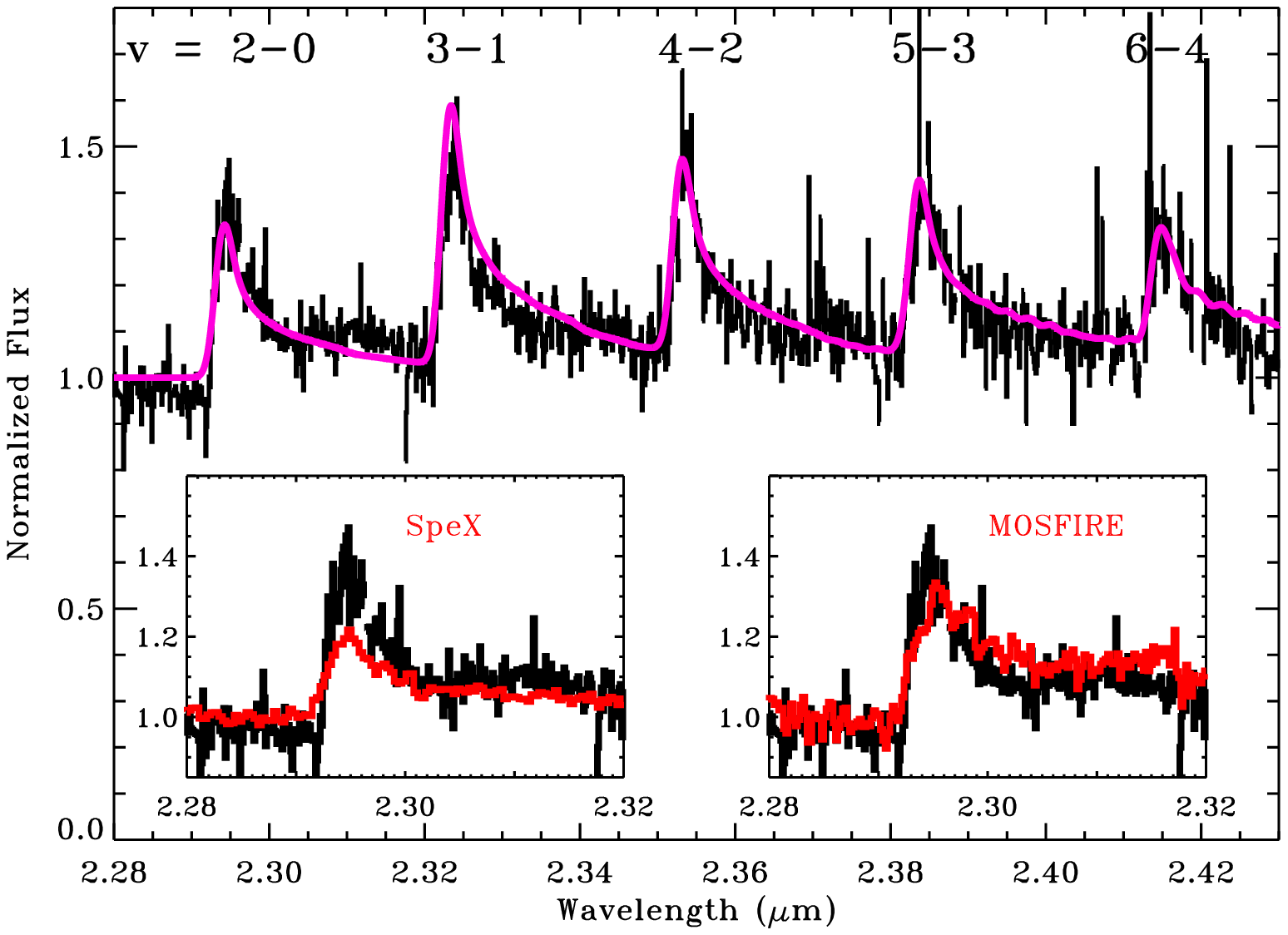}
\caption{The CO overtone emission from ASASSN-15qi during outburst.
  The binned IGRINS spectrum (black in both plots) is well
  reproduced by emission from CO with an excitation temperature of
  4000 K (magenta spectrum).  The CO emission in the MOSFIRE and SpeX spectra (red
  in the insets) has a similar equivalent width as the emission
  detected with IGRINS, indicating that the CO flux has decayed on a
  similar timescale to the $K$-band continuum emission.}
\label{fig:cocomp}
\end{figure}

\subsection{Disk Emission}
\subsubsection{Non-detection of Dust}
The mid-IR SED is consistent with a single temperature blackbody and
no excess dust emission.  The SED fit shown in Figure~\ref{fig:sed}
does not fully recover the near-IR and mid-IR fluxes, but this small
discrepancy could easily be recovered with a 2-temperature fit to the
SED.  The presence of dust cannot be inferred from this small difference.
Adust disk could be masked by
bright emission from a hot gaseous disk; for FUor objects, the excess
dust emission may be detected only longward of $\sim 5$ $\mu$m
\citep{zhu08}.  At longer wavelengths, the 850 $\mu$m flux upper limit corresponds to a maximum dust mass of
$\sim 7$ M$_\odot$.  This limit is too high to provide any useful
constraints on an envelope or disk.

\subsubsection{A Gaseous Disk?}
Bright emission is detected in the CO overtone bands ($\Delta v =2$)
at 2.3 $\mu$m (Figure~\ref{fig:cocomp}).  The presence of CO overtone emission
from young stars often indicates the presence of a warm disk.

We created synthetic CO emission spectra assuming an optically thin
slab in local thermodynamic equilibrium at a single temperature, with CO energy levels and oscillator strengths were
  obtained from \citet{chandra96}.
The band shape and relative strengths are reasonably well reproduced
with a CO excitation temperature of 4000 K and a broad line profile with a full width at half-maximum intensity (FWHM) of
300 \kms\ centered at the stellar radial velocity.   Individual
lines blend together and are not resolved.  The CO
luminosity in the IGRINS spectrum leads to an estimate of $2\times10^{47}$ warm CO molecules.
The band shape and lack of resolved lines in the high resolution 
IGRINS spectrum requires the high velocity broadening.    
The characteristics of the CO overtone emission are consistent with
expectations for emission from a hot gaseous disk.

The band equivalent width is similar in the IGRINS (21 days
post-outburst) and MOSFIRE
(89 days post-outburst) epochs and lower in the early SpeX observation
(13 days post-oubturst).  Because of the decay of the continuum brightness, the total flux and therefore
number of emitting molecules was roughly a factor of $1.7$ higher during the
IGRINS observation than either the SpeX or MOSFIRE observation.  The
CO was not spatially extended beyong the continuum in any observation.

\begin{figure*}[!t]
\epsscale{0.7}
\plotone{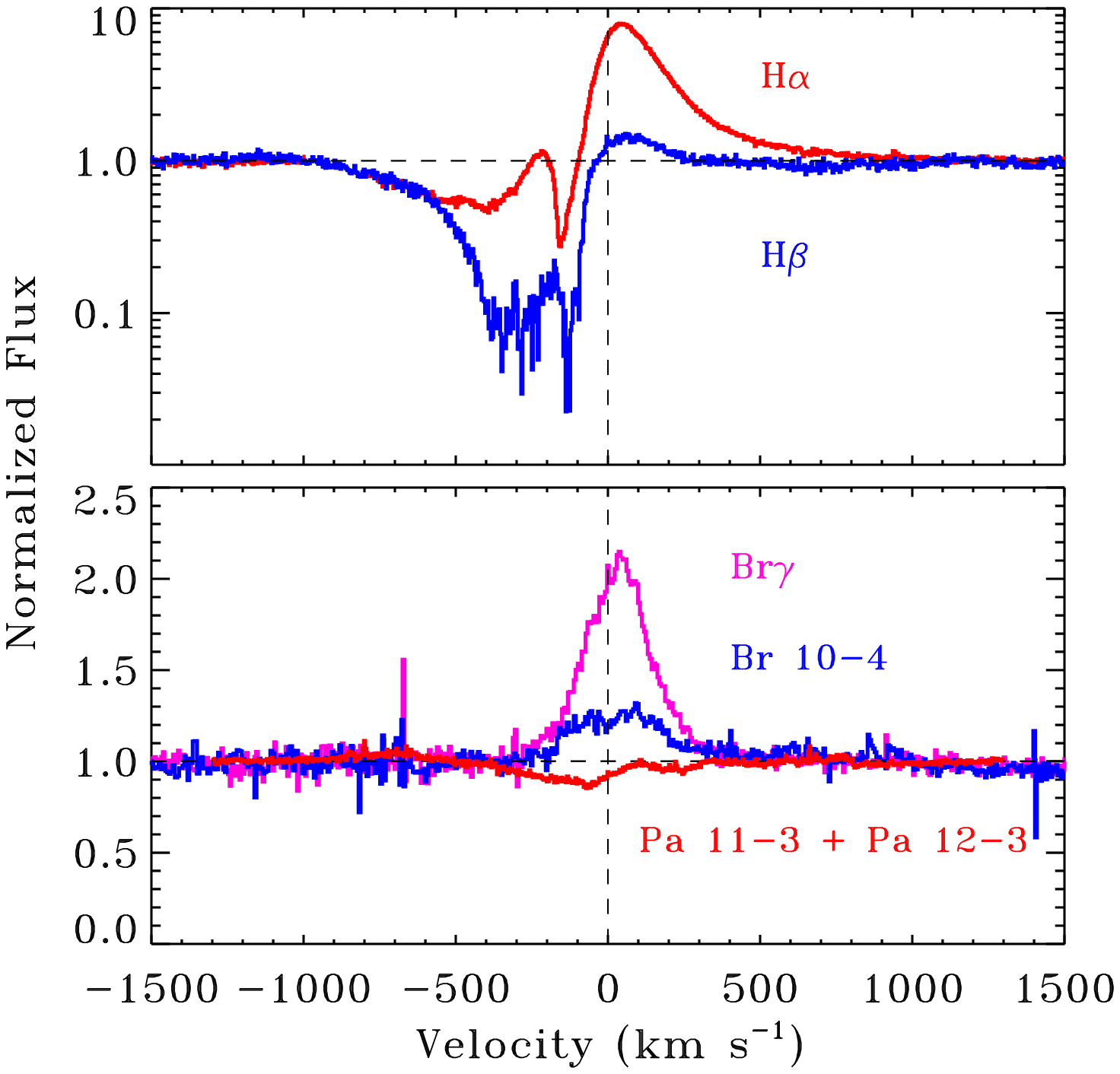}
\epsscale{1.0}
\plottwo{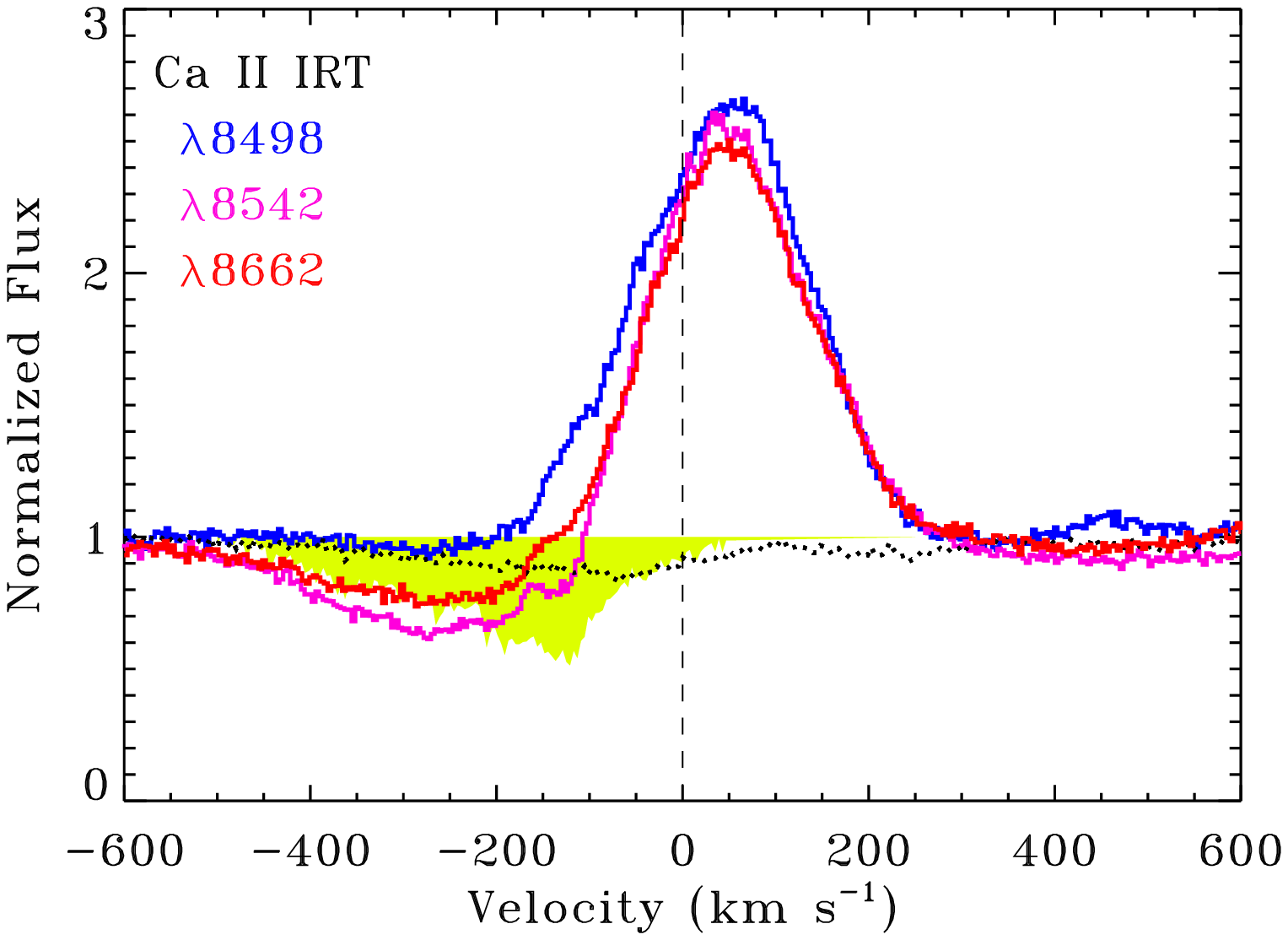}{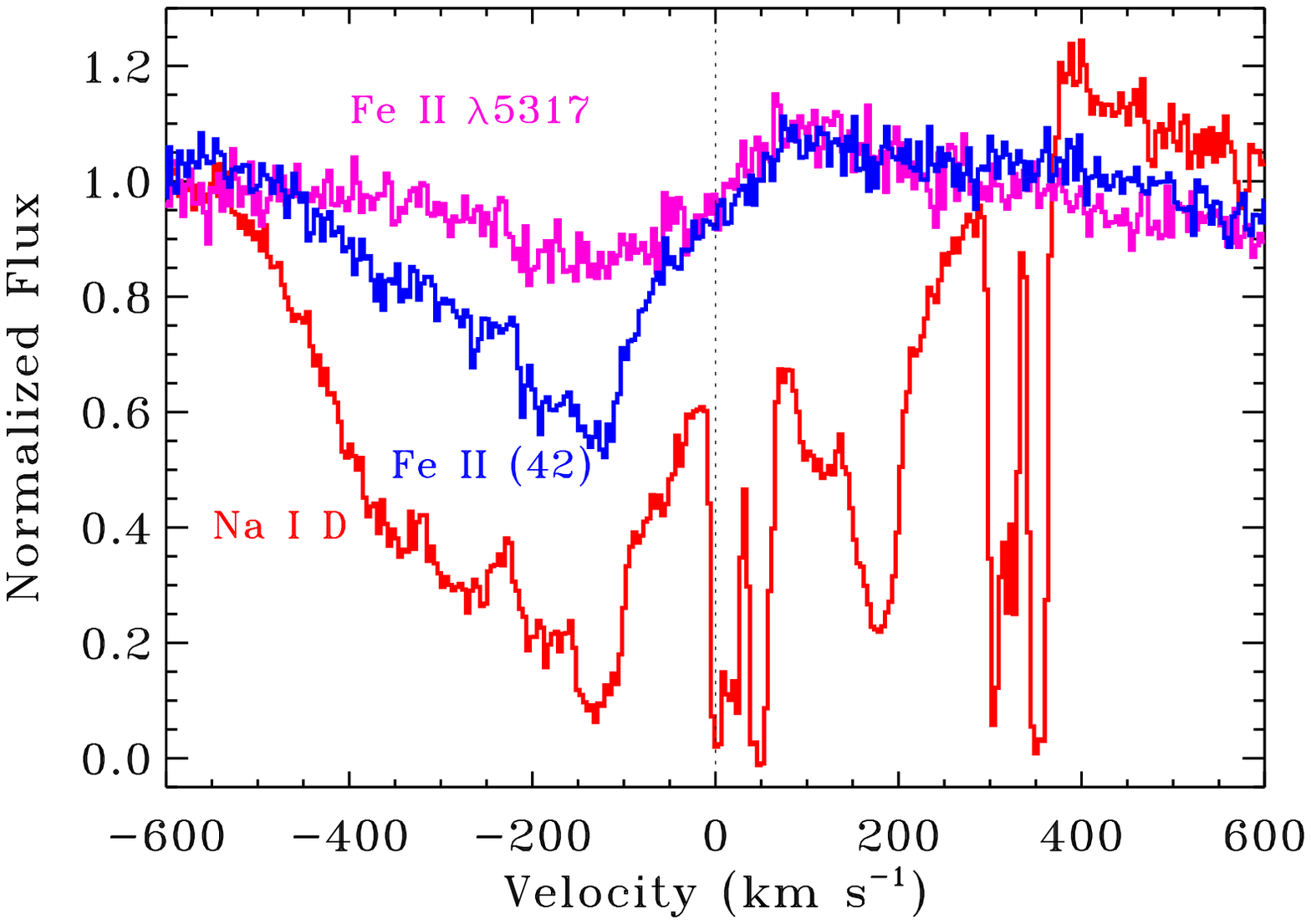}
\caption{Wind absorption features seen in the Keck/HIRES spectrum
  obtained 25 days after the outburst peak.  Top:  High resolution spectra of H lines during the outburst.  The
  wind absorption is detected to $\sim 1000$ \kms\ in H$\alpha$ and
  H$\beta$.  Some scattered emission is detected at the core of the
  wind absorption profile from 100--400 \kms.  The H lines are
  detected in emission up to $n\approx11$ in the Br and Paschen series.
  Several Paschen lines still show weak wind absorption. Bottom left:  
 The \ion{Ca}{2} IR triplet shows strong emission and
  weak, optically thin absorption detected to about $-500$ \kms. The shaded
  yellow region shows \ion{Fe}{2} wind absorption, while the dashed
  line shows the profile of high order Paschen lines.  Bottom right:
  The opacity in the  \ion{Fe}{2} and \ion{Na}{1} D wind
  absorption profiles is maximum at $-140$ \kms\ and is detected to velocities of $-500$ \kms.}
\label{fig:hlines}
\end{figure*}

\subsection{Wind Absorption and Emission during Outburst}

The diversity of wind absorption lines in the optical spectrum of
ASASSN-15qi during outburst is fascinating.  In the following subsections, we discuss the wind
absorption features in different diagnostics.  We then describe the
disappearance of this wind.

\subsubsection{H Lines}

The H lines in ASASSN-15qi all look remarkably different (Figure~\ref{fig:hlines}).  The Balmer lines H$\alpha$ and
H$\beta$ both exhibit strong blueshifted absorption and redshifted
emission.  At velocities of $0$ to $-500$ \kms\ (relative to the
assumed radial velocity of $-65$ \kms), the lines show
absorption that is filled in with some emission, likely from scattered
light in the wind.  The wind scattering also leads to redshifted
emission in H$\alpha$, strong out to 500 \kms\ and detectable to 1000
\kms.  The red wing of
H$\beta$ emission is weak and only detected to $\sim 300$ \kms.  

The H$\beta$ wind
absorption is deep between $-100$ to $-400$ \kms\ and gradually becomes
weaker until the terminal velocity around $-1000$ \kms.  Much of the
wind absorption in H$\alpha$ is contaminated by emission.   However,
the H$\alpha$ and H$\beta$ wind absorption depths appear to match$^7$ between
$-600$ to $-1000$ \kms, with a depth of $\sim 40$\% at $-600$ \kms\
and a few percent at $-1000$ \kms.  The wind absorption in these
lines may be optically thick but covers only a fraction of the
emission region.  Alternately, the wind covering fraction may be
high even at large velocities, but with a low optical depth.
\footnotetext[7]{In the H$\alpha$ line, much of this region is located
  between two orders and falls off the detector.  However, enough of
  the line is detectable that the statement about the similarity of
  H$\alpha$ and H$\beta$ is likely correct.}

High order Paschen lines are detected with shallow blueshifted absorption to $-600$
\kms\ and a photospheric absorption component that extends to about $\pm300$ \kms.
The lower order Paschen lines show
emission and weak absorption in the low resolution spectrum of
\citet{connelley15}.  In contrast, the Brackett lines (Br$\gamma$ and
higher) are seen in emission up to Br 15 (H 15-4).  These emission line 
profiles are similar in shape to the inverse of the Paschen absorption line
profiles.  The lack of emission in higher order H lines suggests that collisions and
some photoexcitation determine the hydrogen level populations.

\subsubsection{He Lines}
\citet{connelley15} detected a classic P Cygni profile in
the \ion{He}{1} $\lambda10830$ line 13 days after the outburst peak.
The line looks similar to the nearby Paschen lines, with absorption seen
out to about $-1000$ \kms, strong emission from $0-500$ \kms, and an
emission wing that extends to about $+1000$ \kms.  The lower level of the \ion{He}{1}
$\lambda10830$ line is meta-stable and may be dominated by scattering
when the opacity in the lower level is sufficiently high
\citep[e.g.,][]{fischer08}.

Optical He lines are not detected in either emission or
absorption.  Emission in optical \ion{He}{1} lines is commonly used as
a diagnostic of magnetospheric accretion
\citep[e.g.,][]{beristain01,alcala14,cauley15}.  The lack of any emission in
these lines indicates an absence of magnetopsheric accretion.  The lack of
absorption in the optical He lines suggests temperatures cooler than 
$\sim 15,000$ K \citep{kwan11}.

\begin{figure*}[!t]
\plotone{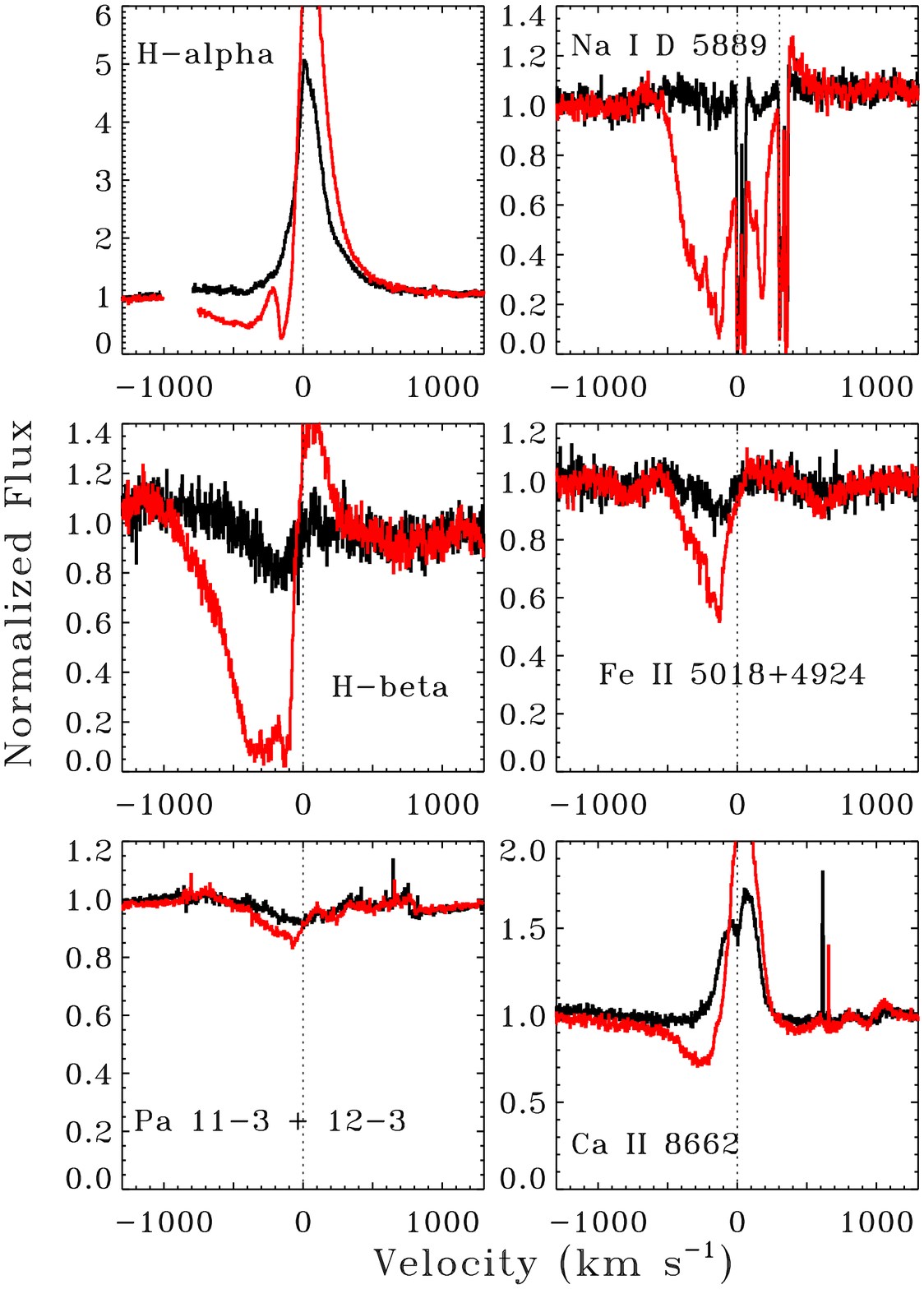}
\caption{A comparison of the two HIRES spectra of ASASSN-15qi.  The deep and fast wind absorption features seen 25 days after
  the outburst peak (red) had disappeared by 80 days after the outburst
  peak (black).}
\label{fig:gonegirl}
\end{figure*}

\begin{table*}
\caption{Line Equivalent Widths}
\label{tab:eqws}
\begin{tabular}{lcccccccc}
\hline
 & Days & H$\alpha$ & H$\beta$ & \ion{Ca}{2} & \ion{O}{1} &
\ion{Na}{1} & \ion{Fe}{2} & \ion{Fe}{2}\\
Spectrum  & post-peak & 6563 & 4861 & 8498 & 7774 & 5893 & 5018 & 4924 \\
\hline
Kanata & 5.5 & $-37\pm0.4$  & $9.0\pm1.4$ & $-5.3\pm0.5$ & $4.0\pm0.3$ & $6.9\pm0.6$ &
-- & --\\
 Bisei & 6.0 & $-34.5\pm0.4$  & $2.6\pm0.3$ & -- & $5.0\pm0.4$ & $5.4\pm0.3$ & $2.25\pm0.09$ &
 $2.46\pm0.10$ \\
 SpeX & 12.9 & -- & -- &$-8.6\pm0.3$ & $3.5\pm0.1$ &-- & -- & --\\
Kast & 17.5 & $-30.2\pm0.2$ & $5.0\pm0.2$ & $-7.5\pm0.2$ & $3.7\pm0.1$ & $7.59\pm0.05$ & $2.24\pm0.07$ & $2.06\pm0.10$
7 \\
 HIRES & 24.7 & $-31.2\pm0.1$ & $7.5\pm0.04$ & $-10.7\pm0.02$ & -- &
 $9.92\pm0.03$& $2.08\pm0.01$ & $1.58\pm0.01$\\
 OSMOS & 40.7 & $-26\pm0.4$ & -- & -- & -- & $4.05\pm0.06$ & $<0.2$ & $<0.2$\\
 HIRES & 79.7 & $-19.65\pm0.04$& $1.70\pm0.02$ & $-5.82\pm0.02$ & --& $1.95\pm0.02$ &
 $0.25\pm0.01$ & $0.67\pm0.01$ \\
 SPRAT & 94.4 & $-29.9\pm0.4$ &-- & -- & -- & $2.9\pm0.3$ & -- & --\\
 SPRAT & 206.4 & $-13.2\pm0.3$ & -- & -- & -- & $2.6\pm0.4$ & -- &
 --\\
LBT & 255.9 & $-8.89\pm0.07$ & $1.0\pm0.1$ & $-2.2\pm0.1$ & $0.78\pm0.02$ & $2.69\pm0.04$ & -- & --\\
\hline
\multicolumn{9}{l}{Measured equivalent widths and 3$\sigma$ upper
  limits, all in units of \AA; negative
  equivalent widths are emission lines.}\\
\end{tabular}
\end{table*}

\subsubsection{Ca II IR Triplet}

The \ion{Ca}{2} IR triplet lines$^8$ have strong emission, with blueshifted
absorption that extends to about $-500$ \kms\ relative to the assumed
radial velocity (Figure~\ref{fig:hlines}).  The blueshifted
absorption is deepest in the $\lambda8542$ line and shallowest in
the $\lambda8498$ line, following the order of oscillator
strengths.  This ordering implies that the wind absorption is
optically thin.  
The emission is much stronger than the absorption.  The \ion{Ca}{2}
$\lambda8498$ line has a centroid velocity of $+40$ \kms\ and a FWHM of 210
\kms, although both values are likely affected by the wind absorption.
\footnotetext[8]{H Paschen lines ($n=13$, 15, 16--3) are located at 100--160 \kms\ to the red of each
of the \ion{Ca}{2} IR triplet lines.  The lower H Paschen lines ($n=12$--3 and 11--3)
have weaker wind absorption than two of the three \ion{Ca}{2} and
are not seen in emission.  These H Paschen lines therefore have a minimal affect
on the \ion{Ca}{2} IR triplet profiles.}

These observed properties are unlike most spectra of young stars.
When the \ion{Ca}{2} line is seen with such large equivalent widths
in other young stars, strong emission is also detected in
\ion{He}{1} lines and is also usually seen in a forest of \ion{Fe}{1} lines.
\citep[e.g.,][]{hamann92,beristain98,gahm08}.
The relative strength of the emission compared with the weak wind
absorption also does not follow
expectations for redshifted emission from wind scattering, as seen in
the H lines and as expected for classic P Cygni profiles.
For young stars with such strong wind absorption, the \ion{Ca}{2} IR
triplet should  show stronger absorption features.

The upper levels of the \ion{Ca}{2} infrared triplet lines are the same upper
levels as the resonant \ion{Ca}{2} H \& K lines \citep[see description
of \ion{Ca}{2} energy levels in, e.g.,][]{li93}.  In a cool wind
exposed to strong UV radiation, the H \& K lines will absorb
radiation.  Radiative de-excitation then occurs through the IR triplet lines,
producing strong emission.  The wind must be sufficiently cool and/or
low density that most of the \ion{Ca}{2} population is in the ground
state, and the central source must be hot enough to provide
enough photons for sufficient \ion{Ca}{2} H \& K absorption to occur
to explain the IR triplet emission.  The lack of [\ion{Ca}{2}]
$\lambda7291$ indicates electron densities $n_e>10^8$ cm$^{-3}$ \citep{nisini05}.

\subsubsection{Other Wind Absorption Features}

Figure~\ref{fig:hlines} shows profiles of the \ion{Na}{1} D lines and
of the coadded
\ion{Fe}{2} $\lambda4924,5018$ lines. Not shown is the \ion{O}{1} $\lambda8446$ line,
which has a profile very similar to \ion{Fe}{2} $\lambda5317$.
The wind absorption in these lines reaches a maximum opacity at 
$-140$ \kms\ and gradually decreases to zero at $-550$ \kms.  At
$-140$ \kms, the flux in H$\beta$ and \ion{Na}{1} is nearly 0, indicating that the wind is
optically thick and covers the entire emission component.  The wind
therefore has an optical depth of $\sim 1$ in \ion{Fe}{2} and several
other lines at $-140$ \kms.
The
\ion{Fe}{2} 42 multiplet lines at 4923, 5018, and 5169 \AA\ are deep and prominent even in
the low resolution spectra.  Some other lines of \ion{Fe}{2} are
weaker but detectable. Wind absorption is weakly detected in
some \ion{Mg}{1} lines and is not detected in \ion{Fe}{1} lines.

\subsubsection{Summary of the Wind Absorption Features}

The wind absorption is prominent in many absorption
lines, with an optical depth that peaks at about $-150$ \kms, and some
gas seen out to $-1000$ \kms.  
The few emission lines that are detected are also likely produced by absorption of
continuum emission by the wind followed by radiative decay.  These
features can all be understood as a fast spherical wind (John Kwan,
private communication).  

The decrease in the wind absorption to faster
velocities likely indicates that the wind is more optically thin at
higher velocities,
although the covering fraction of the continuum emission surface may
also be decreasing.  The deep absorption trough over a wide velocity
range is likely explained by radial acceleration of the wind in our line of
sight.  In radiative transfer models, this acceleration and consequent
deep trough is consistent with expectations for a spherical wind and
inconsistent with narrow absorptions that are produced by disk winds \citep{kwan07,kurosawa11}.

The emission line strengths support the interpretation of the wind as
spherical.  The line emission is likely produced by absorption and re-emission of photons by the
wind into our line of sight.  Most of the emission is seen at $-200$
to $+200$ \kms, consistent with much of the wind having velocities
that are perpendicular to our line of sight.
The strength of the emission indicates
that the wind covers a large fraction of the solid angle centered at
the continuum source.  The properties of the emission lines, together
with the deep absorption troughs, both indicate that the wind is spherical.

\subsection{The Disappearance of the Wind}

The strong, fast wind is the most distinctive feature of the optical
spectrum of the ASASSN-15qi outburst.  This strong wind was
detected with deep absorption features in low resolution spectra and
confirmed as a fast wind in a high resolution spectrum obtained 25
days after peak.  At this phase, the equivalent width of the \ion{Fe}{2} lines had
already started to decline.  In a low resolution spectrum obtained 40 days after peak, the \ion{Fe}{2} lines were
only weakly detected and the \ion{Na}{1} D lines were also weaker than
seen in earlier spectra.

\begin{figure}[!t]
\hspace{-8mm}
\epsscale{1.13}
\plotone{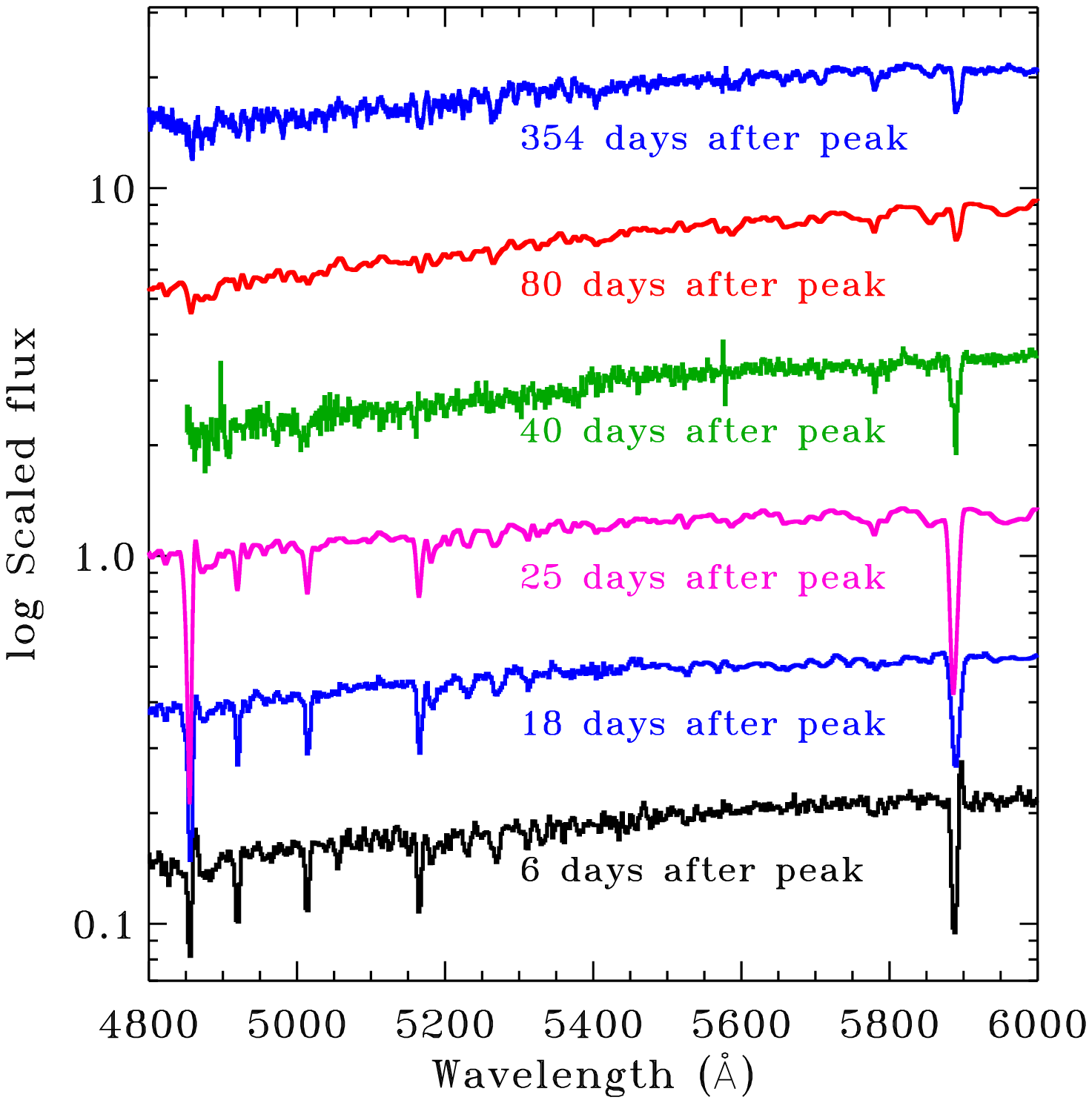}
\caption{Low resolution spectra showing that strong absorption in \ion{Fe}{2}
and \ion{Na}{1} lines starts to disappear 18--40 days after
the outburst peak.}
\label{fig:fe2fade}
\end{figure}

By day 80, our second HIRES spectrum showed that the wind had
mostly disappeared (Figures~\ref{fig:gonegirl}--\ref{fig:fe2fade} and
Table~\ref{tab:eqws}).
  No wind is detectable in any lines at velocities faster than 500 \kms.
From $-100$ to $-500$ \kms, the wind absorption disappeared
entirely from the \ion{Na}{1} D lines, although weak absorption at these
velocities is still
present in the H$\beta$ and \ion{Fe}{2} lines.   The \ion{Ca}{2} wind
absorption also disappeared.  The absorption in the high-order Paschen lines is now
symmetric about the stellar radial velocity and may be photospheric.  
A low resolution spectrum obtained 94 days after the peak was nearly
featureless except for Balmer lines and confirmed that the wind
continued to disappear.  The final spectrum obtained $\sim 250$ days
after outburst no longer shows any wind absorption features.   The
H$\alpha$ line is a symmetric emission profile.

The emission lines also became gradually weaker during the decay, indicating that the wind is also
absorbing fewer photons along other lines of sight.  
 The \ion{Ca}{2} lines line fluxes decrease by a
factor of 3.3 (after adjusting the equivalent width by the continuum
flux) between the first two HIRES epochs (25 and 80 days after peak),
and by another factor of 5 in the last epoch (254 days).  In that
final epoch, the \ion{Ca}{2}
$\lambda8542$ is $\sim 2$ times weaker than the other \ion{Ca}{2}
lines - a very unusual ratio.  The three lines had similar fluxes in all
previous epochs.

\begin{figure}[!t]
\hspace{-8mm}
\epsscale{1.13}
\plotone{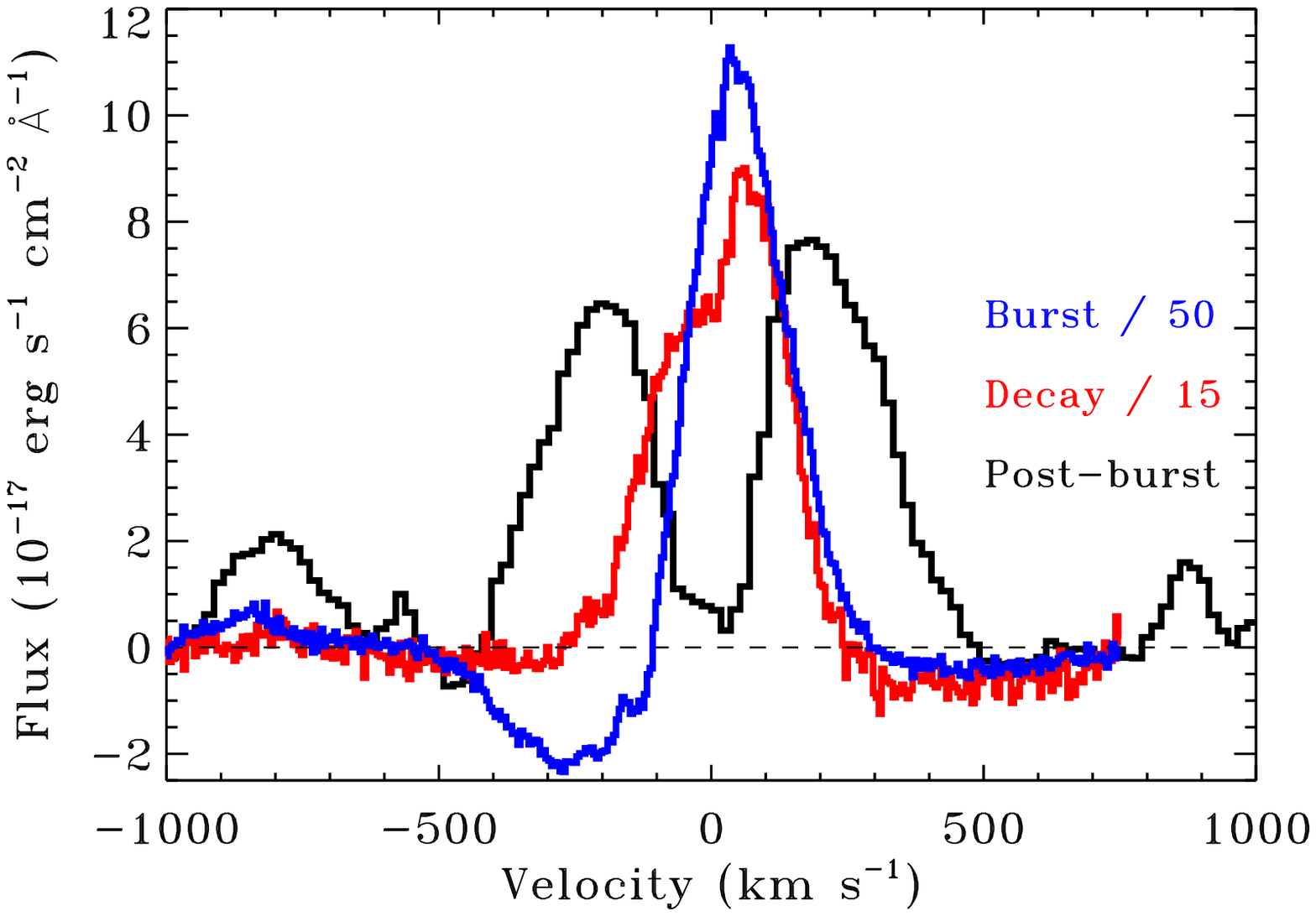}
\caption{\ion{Ca}{2} $\lambda8542$ line during the outburst (blue,
  divided by 50 for visualization),
  decay (red, divided by 15),
  and post-outburst (black).  The double-peaked emission in the
  post-outburst spectrum corresponds to velocity where absorption was
  seen in the first outburst epoch.}
\label{fig:ca2}
\end{figure}

A central absorption component is present in the decay epoch, which
may be related to either self-absorption, the 
photosphere, or lack of emission (see Figure~\ref{fig:ca2}).  By the final post-outburst spectrum, the
lines are double-peaked.  In this last epoch, the \ion{Ca}{2} IRT emission corresponds to
the location of absorption in the first epoch.  The double-peaked
shape of the line profile may imply some directionality 
(e.g., collimation) in the wind.

\section{ATTEMPTING TO CLASSIFY THE 2015 OUTBURST OF ASASSN-15qi}

\subsection{Summary of Observed Properties}

{\it Youth:}   ASASSN-15qi is likely young, as inferred from its
projected location near molecular gas and \ion{H}{2} regions, 
its radial velocity consistent with membership in this parent cloud, 
the projected association with
filaments, and the direct association with nebulosity (see \S 3.1).  
Most of the discussion below assumes that ASASSN-15qi is young.
However, while the star must
be physically located within the molecular cloud, this location could be a
coincidence if ASASSN-15qi is an old object that has migrated into the
molecular cloud.  Such coincidences must be rare but may occur as 
transient surveys
cover larger areas at higher cadence.  The location in a filament
may even be caused by past outflow events that warm nearby dust.

{\it Near-instantaneous Rise Time:}  The
rise time from quiescence to outburst peak is unresolved and $<23$
hours.  The outburst initially decayed by 1 mag over 9 days and another
1.5 mag over 40 days.  The decay then stalled out for 30 days, perhaps
because of the delayed interaction with nebular material, before
resuming its decay and returning to the quiescent state within 200
days (see \S 3).

{\it Fast, cool wind:}  A fast, cool wind was associated with the
outburst and disappeared with the outburst decay.  The
wind during outburst reached 1000 \kms, with a peak optical depth at
150 \kms.  The wind is
likely spherical and not a disk wind.  For a 2 M$_\odot$ star, an escape velocity of 1000 \kms\ corresponds to 0.75 R$_\odot$, which is smaller
than the expected stellar radius (but much larger than the
radius of any compact object).  The peak optical depth in the wind
absorption at $-150$ \kms\ corresponds to a launch radius of $\sim 30$
R$_\odot$ (see \S 3.4-3.5).

{\it Hot photosphere during the outburst:}  The Balmer continuum and optical colors during the
outburst reveal the presence of a $\sim 10,000$ K photosphere associated with
the outburst.  This hot gas disappeared as the outburst faded.  A
strong far-UV radiation field, which may be related to the outburst, is inferred through the presence of
vibrationally excited H$_2$ emission (see \S 3.1--3.2).

{\it Underlying quiescent photosphere:}   The quiescent SED and the
post-outburst spectrum are consistent with an F8 ($6200$ K) star with
$A_V=3.5$ mag.  This photospheric component was detected during the
outburst (see \S 3.2).

{\it Quiescent and Outburst Luminosity:}  The quiescent luminosity is
$\sim 18$ L$_\odot$, which would correspond to an object with
$M\approx2.4$ M$_\odot$ and $6200$ K at 3 Myr, based on the
\citet{feiden16} pre-main sequence evolutionary models.  Assuming no change in extinction, the peak outburst
luminosity is $\sim 1000$ L$_\odot$ (see \S 3.2).     The total
  outburst energy that radiatively escapes is $\sim 7\times10^{42}$ erg over 6 months, with
  33\% released over the first 10 days and 84\% over the first 100
  days. 

{\it A gaseous inner disk but no warm dust:} Strong CO overtone
emission indicates the likely presence of a gaseous disk.  However, no
dust excess is detected in the mid-IR.   The non-detection of sub-mm 
emission is not a significant constraint on the presence of either a
disk or an envelope (see \S 3.3).  The similarity in wind absorption
in H$\alpha$ and H$|beta$ profiles could be explained with an
absorption depth that corresponds to the fraction of emission
intercepted by the wind.

{\it Recurring outburst:}  Photometry from the USNO survey indicates a
likely outburst in 1976 (see \S 3.2).

\vspace{6mm}

\subsection{Possible Causes of the Outburst}

The empirical characterization of the outburst of ASASSN-15qi is
supposed to help to identify
the responsible phenomenon.  The fast rise time and the maximum wind velocity both point
to an event at or very near the star.  The outburst coincided with a
cool wind launched spherically (or nearly so) from the star and/or an
inner disk.  In the following subsubsections, we discuss
several different scenarios that might explain this outburst.

\subsubsection{A YSO Accretion Event (EXor or FUor)?}  

Accretion is the default explanation for an
increased luminosity from a young star.  In this case, a large
accretion rate could lead to additional viscous heating in the inner disk.  The
inner disk would then become hotter and brighter, leading to bluer spectrum.
If the observed luminosity is related to an
accretion event, then the accretion rate at the outburst peak
could be estimated as
\begin{equation}
\dot{M}=\frac{LR_*}{GM_*}\approx 6\times10^{-5} \frac{L_{\rm burst}}{1000 {\rm L}_\odot}
\frac{R_*}{2 {\rm R}_\odot}\frac{M_*}{{\rm M}_\odot}^{-1}  {\rm M}_\odot {\rm yr}^{-1},
\end{equation}
assuming that the luminosity released by viscous heating is directly
related to the increase in accretion energy and that the accretion
energy is efficiently converted into radiation.  

The spectra lack any
indication of magnetospheric accretion; the emission features
can all be explained by wind scattering$^9$.  These magnetospheric
accretion features, including
many strong optical emission lines, would be expected in an EXor
\citep[e.g.,][]{herbig89}.  Although strong CO overtone emission is typical of
EXors and often used to classify outbursts
\citep[e.g.,][]{lorenzetti09}, but given other discrepancies, this diagnostic is
insufficient to diagnose the outburst of ASASSN-15qi.
\footnotetext[9]{When accretion is in a magnetospheric geometry, the H-line luminosities
  may be converted into an accretion luminosity following relations
  established by \citet[e.g.,][]{alcala14}.  However, the likely
  importance of absorption and scattering in generating H emission and
  forming the line profiles, when combined with the lack of optical \ion{He}{1}
  emission, suggests that the H lines of ASASSN-15qi are not produced by magnetospheric
  accretion.}

The fast timescale could be explained by the lack of any magnetospheric cavity
in the inner disk.  Although the leading explanation for EXor events
explains a buildup of mass in the inner disk through a large
magnetospheric cavity \citep{dangelo10}, EXor events could instead be
triggered by disk instabilities at larger radii \citep[see suggestion
by][]{zhang15}.  However, the
challenge for applying this interpretation to ASASSN-15qi is the lack
of excess dust emission in the SED.  

In principle, in an FUor the near-IR dust emission could be masked by
strong emission from a hot, optically thick disk, as seen in the SED models of
\citet{zhu08}.  The quiescent luminosity of $\sim 18$ L$_\odot$
measured from the quiescent spectrum and 
SED would then be interpreted as disk emission.    This luminosity would be
extreme for normal disks but fainter than the $\sim 200$ L$_\odot$
emitted from FUor disks \citep{zhu08}.  
The B-star-like Balmer jump
seen in the ASASSN-15qi
spectrum (see FIg.~\ref{fig:speclo}) is present in other FUor outbursts \citep{chalonge82,szeifert10}.  
The lack of magnetospheric accretion signatures is also consistent with
expectations for FUors, since the accretion rates are high enough to
crush stellar magnetic fields.  . 
However, the 2015 outburst was an eruptive event with a near-instantaneous
rise time and short decay, whereas FUor (and EXor) outbursts are driven by changes in
viscous disks on timescales of months-to-years.  
This challenge could be solved if the 2015
outburst occurred
during a much longer FUor-like outburst.

\subsubsection{A YSO Extinction Event?}   In principle, the increased brightness
could be caused by a decrease in the extinction.  Such a rapid change
in extinction would likely be related to geometry, as seen in some
young stars with disks
\citep[e.g.,][]{cody14}, rather than dust destruction.
However, the spatially extended nebulosity brightens with the
outburst, which demonstrates that the outburst was seen in multiple
directions and not just our line of sight.  An extinction scenario
would require a dust shell that was cleared or destroyed on a $<1$ d
timescale, and then quickly reformed in our line of sight.  Moreover, the interstellar \ion{Na}{1} and DIB absorption did not
change between the two HIRES epochs, obtained 25 and 80 days after the
outburst peak, which implies that the extinction is unchanged.   Therefore, the brightness increase is unlikely an
extinction-clearing event.

\subsubsection{A Grazing Encounter Between the Star and a Planet?}   

Young planetary systems are thought to evolve through
dynamical interactions between the planets.   Hot Jupiters and
super-Earths that have been detected around more
mature stars may arrive into their present positions by
first being scattered into highly eccentric orbits.  In some cases,
these highly eccentric orbits must take the planet into a grazing
orbit through the stellar photosphere.  The acoustic waves driven by
this interaction could drive a cool wind from the stellar photosphere.
The rocky core of the planet might be able to survive passage through
the atmosphere, so that the interaction would recur on an orbital
timescale.  The dynamical timescale for a grazing encounter would be a
few hours, though the energetics may present a challenge.

A related alternative is a periodic mass transfer event between an
unseen companion or ``companion-tesimal'' 
in an eccentric orbit around the
primary star.  Any multiplicity in this young star may still be
unstable, with interactions that could lead to mass transfer and
ejection.

\subsubsection{A YSO Mass Ejection Event?}  

The wind
absorption and related emission are the most prominent spectral
features of the outburst and disappeared as the outburst faded.
The wind likely plays an important role in this outburst. 
A possible explanation may be a spherical (or nearly
spherical) mass ejection event, perhaps driven by
an material accelerated from the stellar surface or by reconnection at
the star-disk interaction region.
An analogous
scenario on the Sun would be a coronal mass ejection, except that
coronal winds are hot while the wind detected here is cool and likely
has a much larger mass flux.

The fast wind velocity also suggests the a launch radius at or very
near the star.
The wind reaches velocities of 1000 \kms, much faster than the
200--500 \kms\ wind typically seen from
Herbig Ae/Be stars \citep[e.g.,][]{hernandez04,cauley14,cauley15}.  The one exception, Z CMa, has a
wind that reaches $-750$ \kms, and is discussed below.  
 
In this scenario, the wind may
form a photospheric surface, or be launched by the eruption that
heated gas to $10,000$ K.  The outburst strength versus time would
be a function of expansion of the $\tau=2/3$ surface, radiative
cooling, and the decrease of density as the ejected material expands.

\subsubsection{A Classical Be Star Analog?}  

The ASASSN-15qi outburst could be related to structural instabilities
that are caused by rapid rotation, similar to those seen on Classical Be stars.
These main sequence (or near main sequence) stars rotate near their breakup
velocities and as a result periodically release material and form
excretion disks \citep[see review by][]{Porter03}.  The outburst of ASASSN-15qi could
have similar physics in the pre-main sequence that classical Be stars
have at later evolutionary stages.

ASASSN-15qi may be a pre-main-sequence star that is still
contracting, with a compact gaseous disk and no evidence for a dusty
primordial disk.
If the disk is unable to break the spin-up of
the protostar during contraction,
the star may rotate at or close to breakup velocity.  The rise and decay times of ASASSN-15qi
are inconsistent with outburst timescales for classical Be stars, but
this discrepancy could be explained by differences in stellar mass and
radius.  The projected rotational velocity ($v \sin i$) of $\sim180$
\kms\ is about half of the break-up velocity of a $2$ M$_\odot$, $2$
$R_\odot$ star.

\subsubsection{An Fe II Dwarf Nova?}  The rapid rise time of the
ASASSN-15qi outburst corresponds reasonably well to that of novae.  The
brightness increase is smaller than expected for a nova but could
be consistent with that seen in a dwarf nova.  While most nova spectra
are dominated by emission lines from ejecta, early-time \ion{Fe}{2}
novae show similar P Cygni absorption profiles as those seen from
ASASSN-15qi \citep[e.g.,][]{williams92,shafter11}.  However, these
\ion{Fe}{2} absorption spectra quickly turn into emission line
spectra.  Based on the spectra, ASASSN-15qi is not a nova or a CV
(Bob Williams and Fred Walter, private communication).  Despite
different triggers, the similarity in absorption lines indicate
similar excitation conditions in the winds of  \ion{Fe}{2} novae and
ASASSN-15qi.

\subsection{Similarities to Other YSO Outbursts}

The 2015 outburst of ASASSN-15qi is not easily classified.  None of
the above possibilities is compelling.  The default classifications
of outbursts of young stars as EXor or FUor accretion events also do not
seem to apply to ASASSN-15qi.  The outburst rise time was $<1$ day,
compared to 1 month for EX Lup and months-to-years for
FUors \citep[e.g.,][]{kospal11}.  The total outburst duration was also $\sim 3$ orders of magnitude shorter than that of FUor
outbursts and is also shorter than EXor outbursts.

\begin{figure}[t]
\hspace{-8mm}
\epsscale{1.22}
\plotone{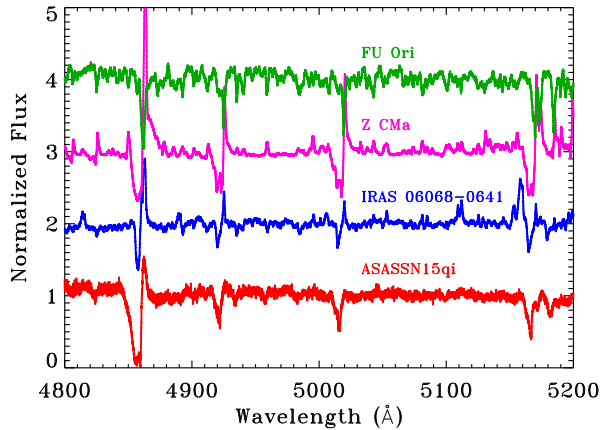}
\caption{A comparison of the high resolution HIRES spectra of ASASSN-15qi
  with a a HIRES spectrum of V899 Mon (obtained in the same setup as ASASSN-15qi), a high resolution CFHT/ESPaDOnS
  spectrum of Z CMa (obtained from the archive at CADC), and a high resolution CFHT/ESPaDOnS spectrum of
  FU Ori (also obtained from the CADC archive and published by Donati
  et al.~2005).  The Z CMa and V899 Mon outbursts have wind absorption
  features that are similar to ASASSN-15qi.  FU Ori has deep, but very
narrow absorption in those same lines and with a much richer
photospheric spectrum than the other objects shown here.}
\label{fig:comparespecs}
\end{figure}

Although these classifications may not apply, the star
shares some characteristics with some previous YSO outbursts.
Figure~\ref{fig:comparespecs} compares the spectrum of ASASSN-15qi with
that of the FUor candidate Z CMa, the outburst of V899 Mon (IRAS 06068-0641), and FU
Ori.  The Z CMa and V899 Mon outbursts have also been
challenging to classify and have similar wind absorption features.
The comparisons in this section are not comprehensive.  Other outbursts, such as
V900 Mon \citep{reipurth12}, have also been difficult to classify, have similar wind
absorption features, and may be similar to ASASSN-15qi but are not discussed below.

The spectrum of the YSO outburst V899
Mon shows wind absorption features  with velocities up to $-700$ \kms\  in many of the same lines as
ASASSN-15qi \citep{ninan15,ninan16}.  The initial discovery
spectrum of the V899 Mon outburst \citep{wils09} showed strong absorption in \ion{Fe}{2},
which disappeared in later spectra \citet{ninan15,ninan16}.  
The \ion{Ca}{2} infrared triplet
lines of V899 Mon look similar to those of ASASSN-15qi, with
redshifted emission and some weak P Cygni absorption.  \ion{He}{1} lines are weak or not
present.

The primary spectroscopic difference is that the CO overtone
bands are seen in absorption in V899 Mon and in emission in
ASASSN-15qi, which may indicate differences in disk structure or in
the surrounding environment.  The V899 Mon outburst was not clearly
identified as either an FUor or EXor outburst.

The outburst timescale is the primary difference between V899 Mon and
ASASSN-15qi.  V899 Mon has had many brightness increases and
decreases \citep{ninan15}.  The initial rise occurred over $>5$ years,
while subsequent fading and brightening episodes occurred over much
shorter timescales (weeks--months).

Another outburst with spectral similarities to ASASSN-15qi is Z CMa,
a binary Herbig Ae/Be system in which one of the two stars 
has been seen in outburst and the other is an emission line object.
The multiple outbursts of Z CMa were classified as
FUor-like when the characteristic CO absorption spectrum was detected
and resolved from the CO emission spectrum produced by the
non-outbursting component \citep{hinkley13}.  
The deep CO absorption indicates the
presence of a bright, viscously heated disk with the warm disk photosphere
located beneath cooler material.  While the CO absorption is one of
the diagnostics for an FUor outburst, the outbursts themselves have
only been $\sim 2$ mag in amplitude with durations of a year, both much smaller
than the classical FUor objects.  

Wind absorption in \ion{Fe}{2} lines is detected in the spectrum of Z
CMa but is only rarely detected to other Herbig Ae/Be stars.  The outflow from Z CMa has a maximum velocity of 750 \kms, a
factor of 2 faster than outflows from other HAeBe stars
\citep{cauley14,cauley15}.

With less data, the outburst of ASASSN-15qi would have been
clearly and uncontroversially described as an EXor.  The primary
characteristics that differ from EXors are the very short rise time,
which is only measurable in surveys having a high cadence, the 
strong wind absorption seen especially in the early-time
optical spectra, and the absence of
spectroscopic signatures of magnetospheric accretion.  The near-IR spectra all show
strong CO emission.  The spectrum of \citet{connelley15} also exhibits strong
emission in many atomic lines.  These spectral features along with an
outburst duration of $\sim 3$ months are amongst the primary
diagnostics for the EXor classification.

\section{SPECULATIONS}

The short outburst of ASASSN-15qi was discovered only because of the
frequent, wide-field monitoring of the ASAS-SN project.  The fast rise
and initial decay left only a $\sim 10$ day period when the outburst
was 3 mag brighter than quiescence.  The frequency of these types of outbursts is highly
uncertain because the detection of months-long bursts requires a high
temporal cadence that are only now becoming available.

The proliferation of all-sky surveys will
require strict criteria for efficient follow-up spectroscopy.  The
rise times of FUor and EXor events, ranging from
a few weeks to 10 years, provide a criterion for selecting FUor and
EXor outbursts for follow-up observations.  However, this choice may also
bias surveys against eruptive outbursts of unknown origin, including those like
ASASSN-15qi.  

The quiescent and outburst SEDs indicate a quiescent
luminosity of $\sim 18$ L$_\odot$, an outburst luminosity of
1000 L$_\odot$, and a total energy radiated of $7\times10^{42}$ erg.  The quiescent SED and the spectroscopic features
during outburst suggest a photospheric temperature of 6000--7000 K,
although the blue/UV excess indicates that a hotter component must also be present.  No warm dust
emission is detected, although a gaseous disk is likely detected in CO
overtone emission and in the shape of the wind absorption profiles.
The energy release accelerated matter spherically from the star to a
maximum velocity of 1000 \kms, as seen in wind emission and absorption
features that accompanied the outburst and disappeared as the outburst faded. 
Archival photometry indicates a previous outburst of ASASSN-15qi in 1976.

These properties cannot be easily explained in the framework of FUor
and EXor outbursts, which are usually applied to large brightness
increases of young stars.  However, other YSO outbursts, including
V899 Mon and Z CMa, share some similar spectrosopic features as
ASASSN-15qi.  Either FUor or EXor events sometimes produce unexpected
physical characteristics, or some protostars
have outbursts that are triggered by different instabilities.  If only
sparse photometry and near-IR spectra were available, this outburst
would have been classified as an EXor.

The outburst timescale and fast wind velocity suggests that the trigger occurred very close
to the star.  Such an event may be
related to mass transfer, interactions between a star and an eccentric
planet, formation of an excretion disk, or by some magnetic
reconnection and outflow event.
We caution readers that we are also unable to rule out with
complete confidence that the outbursts of ASASSN-15qi are produced in the test or
use of advanced alien weaponry.

\pagebreak
\clearpage

\section{Acknowledgements}

We thank the anonymous referee for useful comments and a prompt
report.  GJH thanks John Kwan, Bob Williams, Fred Walter, Dong Lai, Doug Lin, and
Bo Reipurth for interesting conversations about
ASASSN-15qi. We thank LCOGT and its staff for their continued support of ASAS-SN.

GJH is supported by general grant 11473005 awarded by the National
Science Foundation of China.   SD and PC are supported by ``the
Strategic Priority Research Program---The Emergence of Cosmological
Structures'' of the Chinese Academy of Sciences (Grant
No. XDB09000000) and Project 11573003 supported by NSFC. This research
uses data obtained through the Telescope Access Program (TAP), which
is also funded by Grant No. XDB09000000 from the Chinese Academy of
Sciences and by the Special Fund for Astronomy from the Ministry of Finance. 
BS is supported by NASA through Hubble Fellowship grant HF-51348.001
awarded by the Space Telescope Science Institute, which is operated by the
Association of Universities for Research in Astronomy, Inc., for NASA,
under contract NAS 5-26555.  CSK and KZS are supported by NSF grants AST-1515876 and
AST-1515927.  TW-SH is supported by the DOE Computational Science Graduate
Fellowship, grant number DE-FG02-97ER25308.   Support for JLP is in part provided by FONDECYT through grant 1151445
and by the Ministry of Economy, Development, and Tourism's Millennium
Science Initiative through grant IC120009, awarded to The Millennium
Institute of Astrophysics, MAS.
Development
of ASAS-SN has been supported by NSF grant AST-0908816 and CCAPP at the
Ohio State University. ASAS-SN is supported by NSF grant AST-1515927, the
Center for Cosmology and AstroParticle Physics (CCAPP) at OSU, the Mt. Cuba
Astronomical Foundation, George Skestos, and the Robert Martin Ayers
Sciences Fund.  AVF's group at UC Berkeley is grateful for financial assistance from NSF grant
AST-1211916, the TABASGO Foundation, Clark and Sharon Winslow, and the Christopher R. Redlich Fund.
Research at Lick Observatory is partially supported by a generous gift from Google.

This research was made possible through the use of the AAVSO Photometric
All-Sky Survey (APASS) funded by the Robert Martin Ayers Sciences
Fund, data provided by Astrometry.net (Barron et
al. 2008), and filter curves from the Visual Observatory.  The Liverpool Telescope is
operated on the island of La Palma by Liverpool John Moores University in the
Spanish Observatorio del Roque de los Muchachos of the Instituto de Astrofisica
de Canarias with financial support from the UK Science and Technology Facilities
Council.  We thank the {\it Swift} ToO team for responding quickly to
our observation requests.  Some data presented here were made with
the Nordic Optical Telescope, operated by the Nordic Optical Telescope
Scientific Association at the Observatorio del Roque de los Muchachos,
La Palma, Spain, of the Instituto de Astrofisica de Canarias.  Some
data presented here were obtained at the W. M. Keck Observatory, which
is operated as a scientific partnership among the California Institute of
Technology, the University of California, and the National Aeronautics
and Space Administration. Some ESPaDOnS data were downloaded from the CFHT
archive, which is supported by the Canadian Astronomy Data Centre.  The Observatory was made possible by the
generous financial support of the W. M. Keck Foundation.
This work used the Immersion Grating Infrared Spectrograph (IGRINS)
that was developed under a collaboration between the University of
Texas at Austin and the Korea Astronomy and Space Science Institute
(KASI) with the financial support of the US National Science
Foundation under grant AST-1229522, of the University of Texas at
Austin, and of the Korean GMT Project of KASI.
The James Clerk Maxwell Telescope is operated by the East
Asian Observatory on behalf of The National Astronomical Observatory
of Japan, Academia Sinica Institute of Astronomy and Astrophysics, the
Korea Astronomy and Space Science Institute, the National Astronomical
Observatories of China and the Chinese Academy of Sciences (Grant
No. XDB09000000), with additional funding support from the Science and
Technology Facilities Council of the United Kingdom and participating
universities in the United Kingdom and Canada.  The authors
wish to recognize and acknowledge the very significant cultural role
and reverence that the summit of Maunakea has always had within the
indigenous Hawaiian community.  We are most fortunate to have the
opportunity to conduct observations from this mountain.

\section{Affiliations}

\vspace{-6mm}

\footnotetext[1]{Kavli Institute for Astronomy and Astrophysics, Peking University, Yi He Yuan Lu 5, Haidian Qu, 100871 Beijing, People's Republic of China}
\footnotetext[2]{Carnegie Observatories, 813 Santa Barbara Street,
  Pasadena, CA 91101, USA}
\footnotetext[3]{Hubble, Carnegie-Princeton Fellow}
\footnotetext[4]{Department of Astronomy, Peking University, Yi He Yuan Lu 5, Hai Dian District, Beijing 100871, China}
\footnotetext[5]{Caltech, MC 105-24, 1200 E. California Blvd., Pasadena, CA 91125, USA}
\footnotetext[6]{Department of Astronomy, The Ohio State University, 140 West 18th Avenue, Columbus, OH 43210, USA }
\footnotetext[7]{Center for Cosmology and Astro-Particle Physics, The Ohio State University, 191 West Woodruff Avenue, Columbus, OH 43210, USA}
\footnotetext[8]{N{\'u}cleo de Astronom{\'i}a de la Facultad de Ingenier{\'i}a, Universidad Diego Portales, Av. Ej{\'e}rcito 441, Santiago, Chile}
\footnotetext[9]{Millennium Institute of Astrophysics, Santiago,  Chile}
\footnotetext[10]{Department of Astronomy, The University of Texas at Austin, Austin, TX 78712, USA}
\footnotetext[11]{Department of Physics and Astronomy, University of Victoria, Victoria, BC, V8P 1A1, Canada}
\footnotetext[12]{NRC Herzberg Astronomy and Astrophysics, 5071 West Saanich Rd, Victoria, BC, V9E 2E7, Canada}
\footnotetext[13]{Department of Astrophysical Sciences, 4 Ivy Lane, Peyton Hall, Princeton University, Princeton, NJ 08544, USA}
\footnotetext[14]{Key Laboratory for Research in Galaxies and Cosmology, Shanghai Astronomical Observatory, Chinese Academy of Sciences, 80 Nandan Road, Shanghai 200030, China}
\footnotetext[15]{Astrophysics Research Institute, Liverpool Science Park, 146
Brownlow Hill, Liverpool L3 5RF, UK}
\footnotetext[16]{Lunar and Planetary Laboratory, The University of Arizona, Tucson, AZ
85721, USA}
\footnotetext[17]{Earths in Other Solar Systems Team, NASA Nexus for Exoplanet System Science}
\footnotetext[18]{Department of Astronomy, University of California,
  Berkeley, CA  94720-3411, USA }
\footnotetext[19]{ Bisei Astronomical Observatory, 1723-70 Okura,
  Bisei, Ibara, Okayama 714-1411, Japan}
\footnotetext[20]{Coral Towers Observatory, Cairns, Queensland 4870, Australia}
\footnotetext[21]{Institute for Astronomy, University of Hawaii, 640 N. Aohoku Place, Hilo, HI 96720, USA}
\footnotetext[22]{Tuorla Observatory, Department of Physics and Astronomy, University of Turku, V{\"a}is{\"a}l{\"a}ntie 20, FI-21500 Piikki{\"o}, Finland}
\footnotetext[23]{Department of Physical Science, Hiroshima University, 
    1-3-1 Kagamiyama, Higashi-Hiroshima, Hiroshima 739-8526, Japan}
\footnotetext[24]{Hiroshima Astrophysical Science Center, Hiroshima University, 
    1-3-1 Kagamiyama, Higashi-Hiroshima, Hiroshima 739-8526, Japan}
\footnotetext[25]{Okayama Astrophysical Observatory, National Astronomical Observatory of Japan, 3037-5
Honjo, Kamogata, Asakuchi, Okayama, Japan, 719-0232}

\clearpage
\pagebreak

\section{Appendix A:  A Census of H$_2$ Lines}

Table \ref{tab:h2} lists fits to 48 H$_2$ lines detected in the IGRINS
spectrum of ASASSN-15qi.  Lines were identified based on a $>3\sigma$
detection of flux in a narrow (FWHM of 5--10 \kms) line located within $\sim 10$ \kms\ of
the expected line location.  Lines were fit with Gaussian profiles with
flux uncertainties estimated from nearby spectral regions.  
The ability to
detect lines depends on the telluric correction at that wavelength, so additional lines
between 1.4--2.5 $\mu$m may be strong but undetected.

\begin{figure}[!b]
\epsscale{1.1}
\plotone{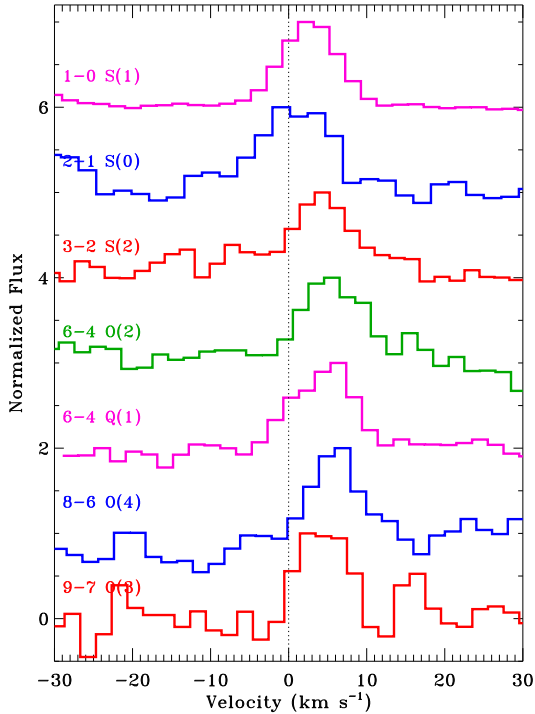}
\caption{Selected profiles of detected H$_2$ lines.}
\label{fig:h2lines}
\end{figure}
 
The strong H$_2$ lines have a median centroid of $-62.5$ \kms\ and a median FWHM
of 8 \kms.  The weak lines are located at $+2$ \kms\ longward of
strong lines, indicating a possible velocity offset between the warm
H$_2$ gas and the fluorescently excited H$_2$ emission.
The standard deviation of 2 \kms\ in the centroid velocities of strong lines is consistent with
expectations for the calibration accuracy of IGRINS.  The equivalent
widths are converted to fluxes from linear fits to the continuum flux,
calculated from $J=12.27$, $H=11.72$, and
$K_S=11.37$ mag (inferred from the near-IR photometry obtained during the outburst).

\begin{table}
\caption{IGRINS H$_2$ Linelist}
\label{tab:h2}
\begin{tabular}{lcccccc}
& $\lambda_{\rm vac}^a$ & Cent.$^b$ & FWHM & EW & $\sigma$(EW) & Flux\\
ID &  ($\mu$m) & \multicolumn{2}{c}{(\kms)}& (\AA) & (\AA)  & $^b$\\
\hline
    3-1   O(4) &       1.4677 &        2.0 &        9.7 &      0.238 &          0.052 &        0.740 \\
   5-3   Q(1) &       1.4929 &        2.8 &       11.3 &      0.204 &          0.039 &        0.620 \\
   4-2   Q(9) &       1.4989 &        3.2 &        7.2 &      0.039 &          0.015 &        0.118 \\
   4-2   O(3) &       1.5099 &        2.6 &       10.8 &      0.146 &          0.016 &        0.436 \\
   5-3   Q(4) &       1.5158 &        3.1 &        5.8 &      0.046 &          0.009 &        0.136 \\
   3-1   O(5) &       1.5220 &        4.1 &        7.8 &      0.072 &          0.015 &        0.211 \\
   5-3   Q(5) &       1.5286 &        3.5 &        9.4 &      0.040 &          0.015 &        0.117 \\
   6-4   S(0) &       1.5369 &       -0.8 &        7.5 &      0.058 &          0.010 &        0.168 \\
   5-3   O(2) &       1.5607 &        4.0 &        8.7 &      0.116 &          0.012 &        0.331 \\
   7-5   S(3) &       1.5615 &        3.2 &        5.3 &      0.026 &          0.007 &        0.074 \\
   4-2   O(4) &       1.5635 &        3.0 &        9.1 &      0.090 &          0.012 &        0.254 \\
   7-5   S(2) &       1.5883 &        5.4 &        8.8 &      0.039 &          0.013 &        0.108 \\
   6-4   Q(1) &       1.6015 &        4.6 &        9.7 &      0.122 &          0.011 &        0.333 \\
   6-4   Q(2) &       1.6074 &        4.2 &        7.9 &      0.096 &          0.009 &        0.260 \\
   5-3   O(3) &       1.6135 &        0.2 &       11.2 &      0.193 &          0.015 &        0.519 \\
   7-5   S(1) &       1.6205 &        8.1 &        5.7 &      0.033 &          0.006 &        0.088 \\
   4-2   O(5) &       1.6223 &        5.7 &        9.2 &      0.064 &          0.011 &        0.172 \\
   6-4   Q(4) &       1.6281 &        4.2 &        7.1 &      0.050 &          0.009 &        0.132 \\
   7-5   S(0) &       1.6585 &        8.4 &        6.5 &      0.040 &          0.011 &        0.103 \\
   5-3   O(4) &       1.6718 &        1.6 &       10.4 &      0.142 &          0.012 &        0.358 \\
   6-4   O(2) &       1.6750 &        5.6 &       10.1 &      0.128 &          0.012 &        0.323 \\
   7-5   Q(1) &       1.7288 &        6.6 &        6.9 &      0.092 &          0.014 &        0.218 \\
   9-7   S(1) &       1.9430 &        4.1 &        6.2 &      0.062 &          0.012 &        0.109 \\
   7-5   O(4) &       1.9434 &        7.7 &        6.3 &      0.131 &          0.015 &        0.230 \\
   8-6   O(2) &       1.9708 &        9.6 &        6.8 &      0.170 &          0.013 &        0.284 \\
   7-5   O(5) &       2.0220 &        4.6 &        7.1 &      0.117 &          0.015 &        0.178 \\
   1-0   S(2) &       2.0338 &        3.1 &        7.9 &      0.694 &          0.014 &        1.035 \\
   8-6   O(3) &       2.0418 &        5.6 &       10.5 &      0.229 &          0.016 &        0.336 \\
   2-1   S(3) &       2.0735 &       -1.9 &        8.5 &      0.327 &          0.014 &        0.451 \\
   9-7   Q(2) &       2.0841 &        1.6 &        8.1 &      0.100 &          0.013 &        0.135 \\
   9-7   Q(3) &       2.1007 &        3.5 &        6.5 &      0.058 &          0.008 &        0.075 \\
   8-6   O(4) &       2.1216 &        6.0 &        8.2 &      0.169 &          0.011 &        0.209 \\
   1-0   S(1) &       2.1218 &        2.7 &        8.3 &      1.234 &          0.025 &        1.531 \\
   3-2   S(4) &       2.1280 &        9.1 &        8.2 &      0.106 &          0.011 &        0.129 \\
   2-1   S(2) &       2.1542 &        1.2 &        8.1 &      0.457 &          0.013 &        0.524 \\
   9-7   O(2) &       2.1727 &        4.6 &        6.1 &      0.093 &          0.008 &        0.101 \\
   3-2   S(3) &       2.2014 &        2.8 &        8.8 &      0.229 &          0.012 &        0.232 \\
   1-0   S(0) &       2.2233 &        1.2 &        8.1 &      1.106 &          0.011 &        1.052 \\
   2-1   S(1) &       2.2477 &        1.2 &        7.9 &      0.772 &          0.020 &        0.681 \\
   9-7   O(3) &       2.2537 &        4.3 &        6.3 &      0.134 &          0.015 &        0.116 \\
   3-2   S(2) &       2.2870 &        4.2 &        9.6 &      0.333 &          0.016 &        0.256 \\
   9-7   O(4) &       2.3455 &        5.8 &        8.9 &      0.260 &          0.031 &        0.157 \\
   2-1   S(0) &       2.3556 &        0.8 &        9.7 &      0.611 &          0.060 &        0.350 \\
   1-0   Q(1) &       2.4066 &        1.9 &        5.0 &      0.540 &          0.025 &        0.231 \\
   1-0   Q(2) &       2.4134 &        2.5 &        7.7 &      1.228 &          0.021 &        0.502 \\
   1-0   Q(3) &       2.4237 &        2.0 &        9.4 &      1.014 &          0.043 &        0.385 \\
  11-9   S(3) &       2.4374 &        8.4 &        8.0 &      0.567 &          0.035 &        0.193 \\
   1-0   Q(4) &       2.4375 &        2.5 &        6.2 &      0.389 &          0.140 &        0.132 \\
\hline
\multicolumn{7}{l}{$^a$Molecular data obtained from \citet{wolniewicz98}.}\\
\multicolumn{7}{l}{$^b$Centroid velocity relative to heliocentric
  velocity of $-65$ \kms.}\\
\multicolumn{7}{l}{$^c$Units $10^{-15}$ erg cm$^{-2}$ s$^{-1}$}\\
\end{tabular}
\end{table}

%\appendix

\pagebreak

\end{CJK*}

\clearpage

\begin{thebibliography}{}


\bibitem[Alcala et al.(2014)]{alcala14}
Alcala, J.M., Natta, A., Manara, C.F., et al.  2014, A\&A, 561, 2

\bibitem[Allard(2014)]{allard14}
Allard, F. in IAU Symp. 299, eds. M. Booth, B. C. Matthews, \& J. R.
Graham, 271

\bibitem[Allen et al.(2012)]{allen12}
Allen, T.S., Gutermuth, R.A., Kryuokova, E., et al.  2012, ApJ, 750, 125

\bibitem[Armitage(2001)]{armitage01}
Armitage, P.J.  2001, MNRAS, 324, 705

\bibitem[Audard et al.(2014)]{audard14}
Audard, M., Abraham, P., Dunham, M.M., et al.  2014, Protostars and Planets VI,
387

\bibitem[Aspin(2010)]{aspin10}
Aspin, C., Reipurth, B., Herczeg, G.J., \& Capak, P.  2010, ApJL, 719, 50

\bibitem[Aspin(2011)]{aspin11}
Aspin, C. 2011, AJ, 141, 196

\bibitem[Azimlu \& Fich(2011)]{azimlu11}
Azimlu, M., \& Fich, M.  2011, AJ, 141, 123


\bibitem[Bae et al.(2014)]{bae14}
Bae, J., Hartmann, L., Zhu, Z., Nelson, R.P.  2014, ApJ, 795, 61

\bibitem[Banzatti et al.(2015)]{banzatti15}
Banzatti, A., Pontoppidan, K.M., Bruderer, S., Muzerolle, J., \&
Meyer, M.R.  2015, ApJL, 798, 16

\bibitem[Barentsen et al.(2014)]{barentsen14}
Barentsen, G., Farnhill, H.J., Drew, J.E., et al.  2014, MNRAS, 444, 3230

\bibitem[Beristain et al.(1998)]{beristain98}
Beristain, G., Edwards, S., \& Kwan, J.  1998, ApJ, 499, 828

\bibitem[Beristain et al.(2001)]{beristain01}
Beristain, G., Edwards, S., \& Kwan, J.  2001, ApJ, 551, 1037

\bibitem[Black \& van Dishoeck(1987)]{black87}
Black, J.H., \& van Dishoeck, E.F.  1987, ApJ, 322, 412

\bibitem[Bonnell \& Bastien(1992)]{bonnell92}
Bonnell, I. \& Bastien, P. 1992, ApJL, 401, 31


\bibitem[Brown et al.(2013)]{brown13}
Brown T. M. et al., 2013, PASP, 125, 1031

\bibitem[Burton et al.(1998)]{burton98}
Burton, M.G., Howe, J.E., Geballe, T.R., \& Brand, P.W.J.L.  1998,
PASA, 15, 194

\bibitem[Caratti o Garatti et al.(2013)]{caratti13}
Caratti o Garatti, A., Garcia-Lopez, R., Weigelt, G., et al.  2013,
A\&A, 554, 66

\bibitem[Cardelli et al.(1989)]{cardelli89}
Cardelli, J.A., Clayton, G.C., \& Mathis, J.S.  1989, ApJ, 345, 245

\bibitem[Cauley \& Johns-Krull(2014)]{cauley14}
Cauley, P.W., \& Johns-Krull, C.M.  2014, ApJ, 797, 112


\bibitem[Cauley \& Johns-Krull(2015)]{cauley15}
Cauley, P.W., \& Johns-Krull, C.M.  2015, ApJ, 810, 5

\bibitem[Ceraski(1906)]{ceraski06}
Ceraski, W.$^6$, 1906, AN, 170, 339;  According to \citep{joy45}, this was the work of
  Lydia Ceraski and not her husband Wladimir.

\bibitem[Chalonge et al.(1982)]{chalonge82}
Chalonge, D., Divan, L., \& Mirzoyan, L.V.  1982, Ap, 18, 161

\bibitem[Chandra et al.(1996)]{chandra96}
Chandra, S., Maheshwari, V. U., \& Sharma, A. K. 1996, A\&AS, 117, 557

\bibitem[Choi et al.(2014)]{choi14}
Choi, Y.K., Hachisuka, K., Reid, M.J., et al.  2014, ApJ, 790, 99

\bibitem[Cody et al.(2014)]{cody14}
Cody, A.M., Stauffer, J., Baglin, A., et al.  2014, AJ, 147, 82

\bibitem[Connelley et al.(2015)]{connelley15}
Connelley, M.S., Reipurth, B., \& Hillenbrand, L.A.  2015, ATel, 8333

\bibitem[Contreras Pena et al.(2016)]{contreras16}
Contreras Pena, C., Lucas, P.W., Kurtev, R., et al.  2016, A\&A, accepted

\bibitem[Crampton et al.(1978)]{crampton78}
Crampton, D., Georgelin, Y.M., Georgelin, Y.P.  1978, A\&A, 66, 1

\bibitem[Cutri et al.(2003)]{cutri03}
Cutri, R.M., Skrutskie, M.F., van Dyk, S., et al.  2003, VizieR Online Data
Catalog, 2246, 0

\bibitem[D'Angelo et al.(2010)]{dangelo10}
D'Angelo, C.R., \& Spruit, H.C.  2010, MNRAS, 406, 1208

\bibitem[D'Angelo et al.(2012)]{dangelo12}
D'Angelo, C.R., \& Spruit, H.C.  2012, MNRAS, 420, 416

\bibitem[Dobashi et al.(2005)]{dobashi05}
Dobashi, K., Uehara, H., Kandori, R., et al. 2005, PASJ, 57, 1

\bibitem[Dobashi(2011)]{dobashi11}
Dobashi, K. 2011, PASJ, 63, 1

\bibitem[Donati et al.(2005)]{Donati05}
Donati, J.-F., Paletou, F., Bouvier, J., \& Ferreira, J.  2005,
Nature, 438, 466

\bibitem[Feiden(2016)]{feiden16}
Feiden, G.  2016


\bibitem[Filippenko(1982)]{filippenko82}
Filippenko, A.V.  1982, PASP, 94, 715


\bibitem[Fischer et al.(2008)]{fischer08}
Fischer, W., Kwan, J., Edwards, S., \& Hillenbrand, L.  2008, ApJ,
687, 1117

\bibitem[Friedman et al.(2011)]{friedman11}
Friedman, S.D., York, D.G., McCall, B.J., et al.  2011, ApJ, 727, 33


\bibitem[Gahm et al.(2008)]{gahm08}
Gahm, G.F., Walter, F.M., Stempels, H.C., Petrov, P.P., \& Herczeg,
G.J.  2008, A\&A, 482, L35

\bibitem[Greene et al.(2008)]{greene08}
Greene, T.P., Aspin, C., \& Reipurth, B.  2008, AJ, 135, 1421

\bibitem[Green et al.(2015)]{green15}
Green, G.M., Schlafly, E.F., Finkbeiner, D.P., et al.  2015, ApJ, 810, 25

\bibitem[Gully-Santiago et al.(2012)]{gully12}
Gully-Santiago, M., Wang, W., Deen, C., \& Jaffe, D.  2012, SPIE,
8450, 2


\bibitem[Hamann(1992)]{hamann92}
Hamann, F.  1992, ApJS, 82, 247

\bibitem[Hartmann \& Kenyon(1996)]{hartmann96}
Hartmann, L., \& Kenyon, S.J.  1996, ARAA, 34, 207

\bibitem[Hartmann et al.(2016)]{hartmann16}
Hartmann, L., Herczeg, G.J., \& Calvet, N.  2016, ARAA, accepted

\bibitem[Henden et al.(2015)]{Henden15}
Henden, A.A., Levine, S., Terrell, D., \& Welch, D.L.  AAS, 336, 16


\bibitem[Herbig(1977)]{herbig77}
Herbig, G.H.~1977, ApJ, 217, 693


\bibitem[Herbig(1989)]{herbig89}
Herbig, G.H.~1989, ESOC, 33, 233

\bibitem[Hernandez et al.(2004)]{hernandez04}
Hernandez, J., Calvet, N., Briceno, C., Hartmann, L., \& Berlind, P.
2004, AJ, 127, 1682

\bibitem[Hillenbrand et al.(2015)]{hillenbrand15}
Hillenbrand, L.A., Reipurth, B., \& Connelley, M.S.  2015, ATel, 8331

\bibitem[Hinkley et al.(2013)]{hinkley13}
Hinkley, S., Hillenbrand, L., Oppenheimer, B.R., et al.  2013, ApJL,
763, 9

\bibitem[Hind(1864)]{hind}
Hind, J.R.  1964, MNRAS, 24, 65

\bibitem[Holoien et al.(2014)]{holoien14}
Holoien, T. W.-S., Prieto, J.L., Stanek, K.Z., et al.  2014, ApJ, 785, 35

\bibitem[Hora et al.(1999)]{hora99}
Hora, J.L., Latter, W.B., \& Deutsch, L.K.  1999, ApJS, 124, 195

\bibitem[Joy et al.(1945)]{joy45}
Joy, A. 1945, ApJ, 102, 168


\bibitem[Kospal et al.(2011)]{kospal11}
Kospal, A., Abraham, P., Goto, M., et al. 2011, ApJ, 736, 72


\bibitem[Kun et al.(2008)]{kun08}
Kun, M., Kiss, Z.T., \& Balog, Z.  2008, in Handbook of Star Forming Regions,
Vol. 1, ed. B. Reipurth (San Francisco, CA: ASP), 136

\bibitem[Kurosawa et al.(2011)]{kurosawa11}
Kurosawa, R., Romanova, M.M., \& Harries, T.J.  2011, MNRAS, 416, 2623


\bibitem[Kwan et al.(2007)]{kwan07}
Kwan, J., Edwards, S., \& Fischer, W.  2007, ApJ, 657, 897


\bibitem[Kwan \& Fischer(2011)]{kwan11}
Kwan, J., \& Fischer, W.  2011, MNRAS, 411, 2383


\bibitem[Laher et al.(2014)]{laher14}
Laher, R.R., Surace, J., Grillmair, C.J., et al.  2014, PASP, 126, 674

\bibitem[Lang et al.(2010)]{lang10}
Lang, D., Hogg, D.W., Mierle, K., Blanton, M., \& Roweis, S.  2010,
AJ, 139, 1782


\bibitem[Law et al.(2009)]{law09}
Law, N.M., Kurlkarni, S.R., Dekany, R.G., et al.  2009, PASP, 121, 886


\bibitem[Li \& McCray(1993)]{li93}
Li, H., \& McCray, R.  1993, ApJ, 405, 730


\bibitem[Lorenzetti et al.(2009)]{lorenzetti09}
Lorenzetti, D., Larionov, V.M., Giannini, T., et al.  2009, ApJ, 693, L1056

\bibitem[Lorenzetti et al.(2012)]{lorenzetti12}
Lorenzetti, D., Antoniucci, S., Giannini, T., et al.  2012, ApJ, 749, 188

\bibitem[Mairs et al.(2015)]{mairs15}
Mairs, S., Johnstone, D., Kirk, H., et al.  2015, MNRAS, 454, 2557

\bibitem[Makovoz et al.(2005)]{makavoz2005}
Makovoz, D., \& Marleau, F.R.  2005, PASP, 117, 1113

\bibitem[Maehara et al.(2015)]{maehara15}
Maehara, H., Ayani, K., Ito, R., Takata, K., \& Kawabata, K.S.  2015,
ATel, 8147


\bibitem[McLean et al.(2012)]{McLean2012}
McLean, I.S., Steidel, C.C. et al. 2012, SPIE Procs, 8446 17

\bibitem[Meyer et al.(2001)]{meyer01}
Meyer, D.M., Lauroesch, J.T., Sofia, U.J., Draine, B.T., \& Bertoldi,
F.  2001, ApJ, 553, 59

\bibitem[Miller et al.(2011)]{miller11}
Miller, A.A., Hillenbrand, L.A., Covey, K.R., et al.  2011, ApJ, 730, 80

\bibitem[Miller \& Stone (1993)]{miller93}
Miller, J.S., Stone R.P.S. 1993, Lick Obs. Tech. Rep. 66
(Santa Cruz: Lick Obs.)

\bibitem[Monet et al.(1998)]{monet98}
Monet, D.G.,et al., 1998, VizieR Online Data
Catalog, 1252, 0

\bibitem[Monet et al.(2003)]{monet03}
Monet, D.G., Levine, S.E., Canzian, B., et al.  2003, AJ, 125, 984


\bibitem[Ninan et al.(2015)]{ninan15}
Ninan, J.P., Ojha, D.K., Baug, T., et al.  2015, ApJ, 815, 4

\bibitem[Ninan et al.(2016)]{ninan16}
Ninan, J.P., Ojha, D.K., \& Philip, N.S.  2016, ApJ, accepted.  arXiv:1605.08533


\bibitem[Nisini et al.(2005)]{nisini05}
Nisini, B., Bacciotti, F., Giannini, T., et al.  2005, A\&A, 441, 159

\bibitem[Park et al.(2014)]{park14}
Park, C., Jaffe, D.T., Yuk, I.-S., et al.  2014, SPIE, 9147, 1

\bibitem[Pickles(1998)]{pickles98}
Pickles, A.J. 1998, PASP, 110, 863


\bibitem[Poole et al.(2008)]{poole08}
Poole, T.S., Breeveld, A.A., Page, M.J., et al.  2008, MNRAS, 383, 627

\bibitem[Porter \& Rivinius(2003)]{Porter03}
Porter, J.M., \& Rivinius, T.  2003, PASP, 115, 1153


\bibitem[Reipurth \& Aspin(2010)]{reipurth10}
Reipurth, B., \& Aspin, C.  2010, in: H. A. Harutyunian, A. M. Mickaelian, Y.
Terzian, eds., Evolution of Cosmic Objects through their Physical Activity,
Yerevan, Gitutyun, p. 19

\bibitem[Reipurth et al.(2012)]{reipurth12}
Reipurth, B., Asplin, C., \& Herbig, G.H.  2012, ApJ, 748, 5

\bibitem[Roming et al.(2005)]{roming05}
Roming P. W. A. et al., 2005, Space Sci. Rev., 120, 95


\bibitem[Shafter et al.(2011)]{shafter11}
Shafter, A.W., Darnley, M.J., Hornoch, K., et al.  2011, ApJ, 734, 12

\bibitem[Shappee et al.(2014)]{shappee14}
Shappee, B..J., Prieto, J.L., Grupe, D., et al.  2014, ApJ, 788, 48

\bibitem[Sicilia-Aguilar et al.(2015)]{sicilia15}
Sicilia-Aguilar, A., Fang, M., Roccatagliata, V., et al.  2015, A\&A,
580, 82


\bibitem[Sreenilayam et al.(2014)]{sreenilayam14}
Sreenilayam, G., Fich, M., Ade, P., et al.  2014, AJ, 147, 53


\bibitem[Stecklum et al.(2015a)]{stecklum15a}
Stecklum, B., Eisloeffel, J., \& Scholz, A.  2015, ATel, 8210


\bibitem[Stecklum et al.(2015b)]{stecklum15b}
Stecklum, B., Eisloeffel, J., \& Wiersema, K.  2015, ATel, 8365

\bibitem[Szeifert et al.(2010)]{szeifert10}
Szeifert, T., Hubrig, S., Sch{\"o}ller, M., Sch{\"u}tz, O., Stelzer,
B., \& Mikulasek, Z.  2010, A\&A, 509, 7

\bibitem[Tody(1993)]{tody93}
Tody, D.  1993, ASP Conference Series 52, 173


\bibitem[Vogt et al.(1994)]{vogt94}
Vogt, S.S. et al., 1994, Proc. SPIE, 2198, 362


\bibitem[Vorobyov \& Basu(2005)]{vorobyov05}
Vorobyov, E.I., \& Basu, S.  2005, ApJ, 633, 137

\bibitem[Williams(1992)]{williams92}
Williams, R.E.  1992, AJ, 104, 725

\bibitem[Wils et al.(2009)]{wils09}
Wils, P., Greaves, J., Catelan, M., et al.  2009, ATel, 2307

\bibitem[Wolniewicz et al.(1998)]{wolniewicz98}
Wolniewicz, L., Simbotin, I., \& Dalgarno, A.  1998, ApJS, 115, 293

\bibitem[Zhang et al.(2015)]{zhang15}
Zhang, K., Crockett, N., Salyk, C., et al.  2015, ApJ, 805, 55

\bibitem[Zhu et al.(2008)]{zhu08}
Zhu, Z., Hartmann, L., \& Calvet, N., et al.  2008, ApJ, 684, 1281

\bibitem[Zhu et al.(2009)]{zhu09}
Zhu, Z., Hartmann, L., \& Gammie, C.  2009, ApJ, 694, 1045


\end{thebibliography}
\end{document}